\documentclass[aps,prb,twocolumn,showpacs,psfig,superscriptaddress,longbibliography]{revtex4-1}
\usepackage{amsfonts}
\usepackage{mathrsfs}
\usepackage{amsmath}
\usepackage{color}
\usepackage{natbib}
\usepackage{textcomp}
\usepackage{graphicx}
\usepackage{bm}
\usepackage{amssymb}

\usepackage{xspace}
\usepackage{epstopdf}
\usepackage{dcolumn}
\usepackage{longtable}
\usepackage{multirow}
\usepackage[colorlinks=true, letterpaper=true, pdfstartview=FitV, linkcolor=blue, citecolor=blue, urlcolor=blue]{hyperref}
\usepackage{float}

\makeatletter

\newcommand{\Rmnum}[1]{\expandafter\@slowromancap\romannumeral #1@}
\makeatother

\begin{document}

\title{Superconductivity in Sodium-Doped T-carbon}

 \author{Jing-Yang You}
 \affiliation{School of Physical Sciences, University of Chinese Academy of Sciences, Beijing 100049, China}

 \author{Bo Gu}
 \email{gubo@ucas.ac.cn}
 \affiliation{Kavli Institute for Theoretical Sciences, and CAS Center for Excellence in Topological Quantum Computation, University of Chinese Academy of Sciences, Beijng 100190, China}
\affiliation{Physical Science Laboratory, Huairou National Comprehensive Science Center, Beijing 101400, China}

 \author{Gang Su}
 \email{gsu@ucas.ac.cn}
 \affiliation{School of Physical Sciences, University of Chinese Academy of Sciences, Beijing 100049, China}
 \affiliation{Kavli Institute for Theoretical Sciences, and CAS Center for Excellence in Topological Quantum Computation, University of Chinese Academy of Sciences, Beijng 100190, China}
 \affiliation{Physical Science Laboratory, Huairou National Comprehensive Science Center, Beijing 101400, China}

\begin{abstract}
T-carbon has been proposed as a new carbon allotrope in 2011, which was successfully synthesized in recent experiments. Because of its fluffy structure, several kinds of atoms can be intercalated into T-carbon, making it a potential versatile candidate in various applications such as hydrogen storage, solar cells, lithium ion batteries, thermoelectrics, photocatalyst, etc. Here we show that superconductivity can appear in Na-doped T-carbon with superconducting transition temperature Tc of 11 K at ambient pressure, and Tc can be enhanced to 19 K under pressure of 14 GPa. The calculations on temperature dependence of specific heat and electrical and thermal conductivities show that the normal state of the Na-doped T-carbon superconductor is a non-Fermi liquid at temperature below 50 K, where the Wiedemann-Franz law is remarkably violated. The prediction of superconductivity in Na-doped T-carbon would spur great interest both experimentally and theoretically to explore novel carbon-based superconductors.

\end{abstract}
\pacs{}
\maketitle


\section{Introduction}
The observation of superconductivity in carbon materials has long been a crucial and fascinating topic in condensed matter physics and materials science, which receives much attention in recent years. Several carbon compounds were reported to be superconductors. For instance, graphite intercalation compounds such as KC$_8$ are known to be superconducting with very low transition temperature Tc~\cite{Lauginie1980}; the boron-doped diamond is a bulk, type-II superconductor below about 4 K~\cite{Ekimov2004}. The discovery of fullerenes C$_{60}$ and C$_{70}$ opens a door to explore new allotropes of carbon, which leads to subsequent flourishing explorations of carbon allotropes~\cite{Kroto1985}.
C$_{60}$ molecules can form solids~\cite{Kraetschmer1990}, which have so large interstitial site spacings that can accommodate intercalants like alkali-metals~\cite{Haddon1991,Hebard1991,Hebard1992}, alkaline earth metals~\cite{Kortan1992,Kortan1992a} and rare-earth elements~\cite{Oezdas1995,Yoshikawa1995}. The superconductivity in alkali-metal doped C$_{60}$ was first discovered in K$_3$C$_{60}$ with Tc of 18 K, and then in Rb$_3$C$_{60}$ with Tc of about 30 K~\cite{Rosseinsky1991,Holczer1991}.
Since then higher Tc of 33 K~\cite{Murphy1992,Tanigaki1991,Tanigaki1993} at 1 bar in Cs$_x$Rb$_y$C$_{60}$ and 40 K in Cs$_3$C$_{60}$ under pressure of 15 kbar~\cite{Palstra1995} were reported. In addition, low temperature superconductivity were also found in pure carbon such as single-walled carbon nanotubes~\cite{Tang2001}, and very recently, the magic twisted bilayer graphene~\cite{Cao2018,Cao2018a}.

A novel carbon allotrope coined as T-carbon was proposed in 2011~\cite{Sheng2011}, which has been successfully synthesized in two independent experiments recently. By applying picosecond pulsed-laser irradiation to a multiwalled carbon nanotube suspension in methanol, T-carbon nanowires have been produced through pseudotopotactic conversion~\cite{Zhang2017}. By making use of plasma enhanced chemical vapor deposition on the substrates of polycrystalline diamond or single crystalline diamond, T-carbon has also been obtained in laboratory~\cite{Xu2020}, implying that massive production of T-carbon would be feasible. From the structural point of view, T-carbon has large interspaces between carbon atoms, leading to its density (1.50 g/cm$^3$) lower than graphite and diamond. The low density with large interspaces between carbon atoms provides broad potential applications in various fields~\cite{Qin2019}, which have attracted much interest to explore the versatile properties of T-carbon. To name but a few, T-carbon nanowires were shown to exhibit mechanical anisotropy and excellent ductility~\cite{Bai2018}; the mechanical properties of T-carbon could be modulated by the strain rate and grain size~\cite{Wang2019}; T-carbon has the lowest lattice thermal conductivity among three-dimensional carbon allotropes and may be used as a thermal insulation material~\cite{Yue2017}; by doping elements, the band gap of T-carbon could be adjusted~\cite{Ren2019,Alborznia2019}; the electron mobility in T-carbon was exposed to be higher than those in conventional electron transport materials such as TiO$_{2}$, ZnO and SnO$_{2}$,
implying its potential for good photocatalysts and solar cells~\cite{Ren2019,Sun2019,Alborznia2019}; the Seebeck coefficient of T-carbon was shown to be comparable with or even larger than those of some excellent thermoelectric materials, indicating its potential as a thermoelectric material for energy recovery and conversion~\cite{Qin2019}; the transport properties of T-carbon can also be modified through applying strain, doping appropriate elements, or cutting it into lower dimensional structures~\cite{Gharsallah2016,Qin2016}; etc.

In this paper, we show that the Na-doped T-carbon can be a novel superconductor with Tc of 11 K at ambient pressure by means of the first-principles calculations, and Tc can reach about 19 K at pressure of 14 GPa. The increase of Tc under pressure for the Na-doped T-carbon was revealed from an enhancement of the electron-phonon coupling due to the shift of the phonon spectral weight to lower frequencies. The superconductivity in Na-doped T-carbon is probably induced by the electron-phonon interaction through Bardeen-Cooper-Schrieffer (BCS) mechanism. It is also found that below 50 K its normal state shows a non-Fermi liquid behavior, where the Wiedemann-Franz law is remarkably violated. This work enriches the family of carbon-based superconductors (CBS) and would spur more investigations on physical properties of CBS.


\section{Calculational method}
Our first-principles calculations were based on the density-functional theory (DFT) as implemented in the QUANTUM-ESPRESSO package~\cite{Giannozzi2009}, using the projector augmented wave method~\cite{Bloechl1994}. The generalized gradient approximation (GGA) with Perdew-Burke-Ernzerhof~\cite{Perdew1996} realization was adopted for the exchange-correlation functional. To warrant an energy convergence of less than 1 meV per atom, the plane-waves kinetic-energy cutoff was set as 80 Ry and the energy cutoff for charge density was set as 1000 Ry. The structural optimization was performed until the forces on atoms were less than 1 meV/\AA. An unshifted Brillouin zone (BZ) k-point mesh of 24$\times$24$\times$24 was utilized for electronic charge density calculations. The phonon modes are computed within density-functional perturbation theory~\cite{Baroni2001} on a 6$\times$6$\times$6 $q$ mesh. The electronic transport properties were calculated with the package BoltzTrap~\cite{Madsen2006}.

\section{Structure and spectra of Na-doped T-carbon}

\begin{figure}[!htbp]
  \centering
  \includegraphics[scale=0.41,angle=0]{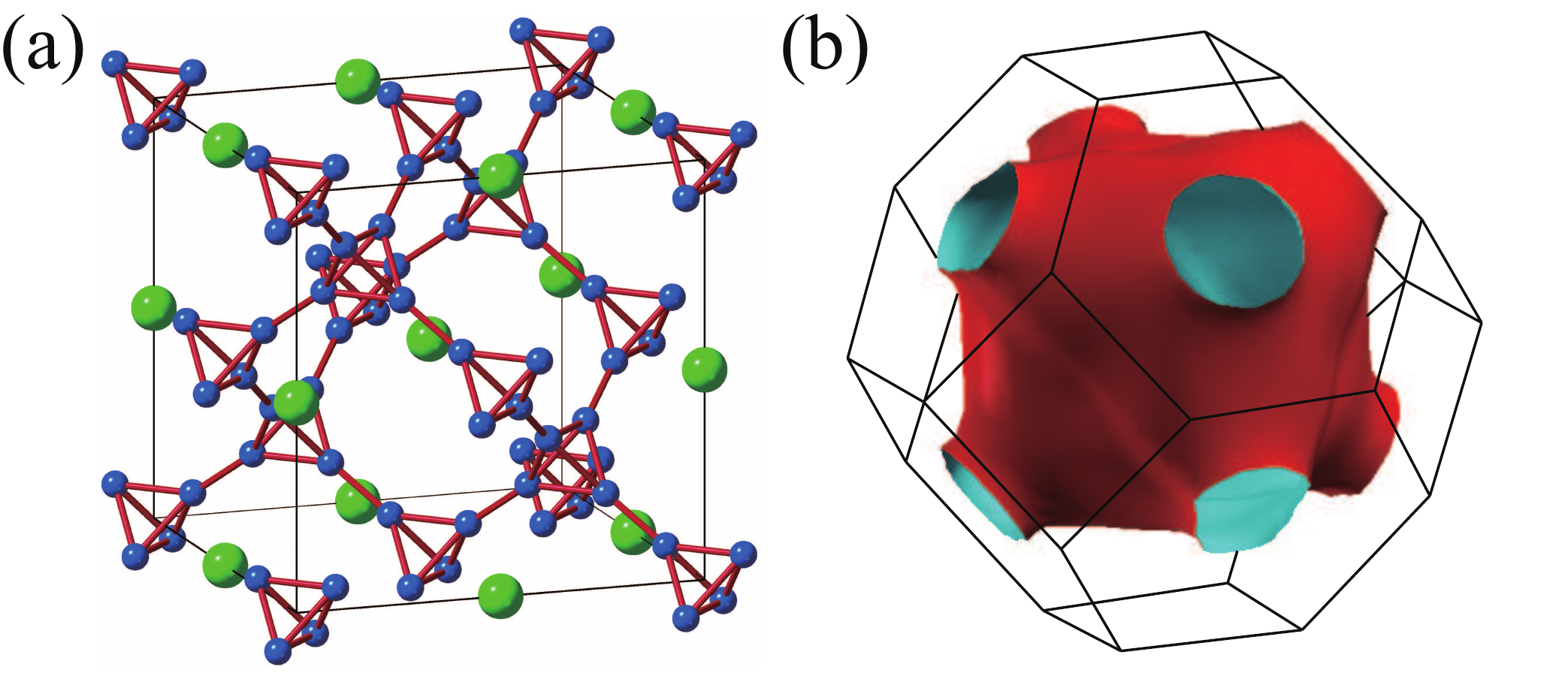}\\
  \caption{(a) The cubic crystalline structure of Na-doped T-carbon NaC$_8$ with Na atoms (green balls) occupying the Wyckoff position $4a$. (b) The Fermi surface of Na-doped T-carbon.}\label{fig1}
\end{figure}

T-carbon possesses a cubic lattice with space group of $Fd\bar{3}m$ (No.227). Each unit cell contains two tetrahedrons with eight carbon atoms, and the lattice constant is about 7.52 \AA. The three unit vectors are $\vec{a}=(l/2)(0,1,1)$, $\vec{b}=(l/2)(1,0,1)$, and $\vec{c}=(l/2)(1,1,0)$, and carbon atoms occupy the Wyckoff position $32e(x; x; x)$ with $x\sim0.0706$. When Na atoms are intercalated into T-carbon, one may obtain the structure of Na-doped T-carbon that possesses the space group of $F\bar{4}3m$ (No.216), where Na atoms occupy the Wyckoff position $4a(0.5; 0; 0)$ and carbon atoms occupy the Wyckoff positions $16e(0.92954; 0.57046; 0.42954)$ or $16e(0.32066; 0.17934; 0.32066)$, as shown in Fig.~\ref{fig1}(a). The optimized lattice constant of Na-doped T-carbon is about 7.5794 \AA, which is slightly larger than that of T-carbon. The total energy calculations show that the optimized structure of Na-doped T-carbon is energetically stable.

\begin{figure}[!htbp]
  \centering
  \includegraphics[scale=0.7,angle=0]{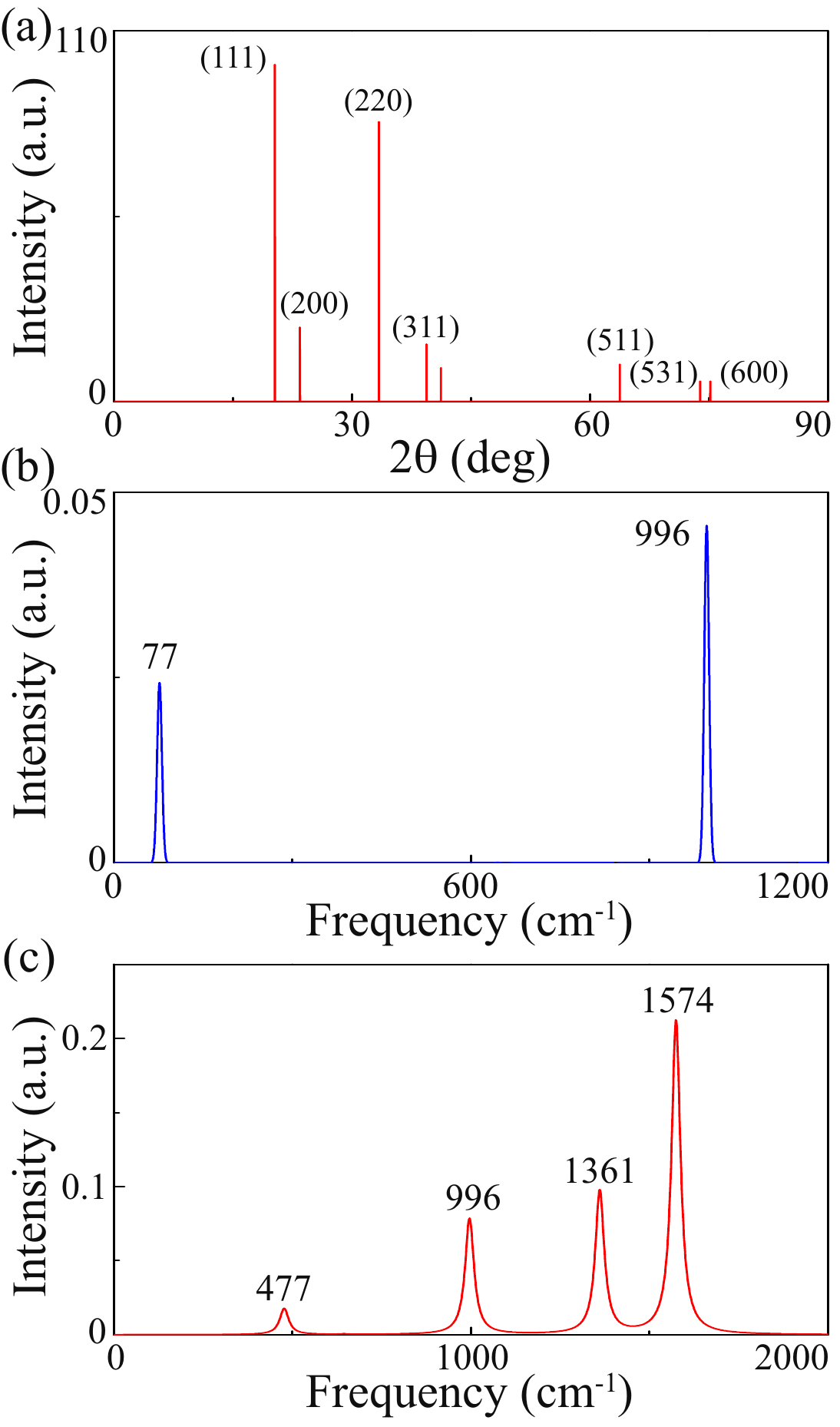}\\
  \caption{(a) The simulated x-ray diffraction (XRD) spectra, (b) infrared (IR) spectra and (c) Raman spectra of Na-doped T-carbon. The X-ray with wavelength of 1.54 \AA \ is used.}\label{fig2}
\end{figure}

To provide more information for possible experimental identifications, we simulated the x-ray diffraction (XRD) spectra of Na-doped T-carbon with wavelength 1.54 \AA. The results are presented in Fig.~\ref{fig2}(a). The XRD spectral peaks appear at the angles 2$\theta$=20.3$^\circ$ of (111), 23.5$^\circ$ of (200), 33.4$^\circ$ of (220), 39.4$^\circ$ of (311), 63.8$^\circ$ of (511), 73.9$^\circ$ of (531) and 75.2$^\circ$ of (600). The simulated infrared (IR) and Raman vibrational modes with corresponding frequencies are presented in Figs.~\ref{fig2}(b) and (c), respectively. The IR spectra show two peaks at 77 cm$^{-1}$ (2.31 THz) and 996 cm$^{-1}$ (29.90 THz). The Raman spectra exhibit well-marked peaks at 477, 996, 1361 and 1574 cm$^{-1}$. These attainable features may be useful for future experimental identification of Na-doped T-carbon.

In addition to the stable structure of Na-doped T-carbon [NaC$_8$ in Fig.~\ref{fig1}(a)], we have also tried to intercalate other alkali metals into T-carbon, and found that the so-obtained structures are either unstable or collapsed. When intercalating Li into T-carbon, we found the structure is not stable dynamically because of its imaginary phonon modes. The other alkali metals such as K, Rb and Cs atoms have so large atom radii that the structure will collapse when intercalating these atoms into T-carbon. Within alkali metals, we uncovered that Na atoms can be intercalated into T-carbon only in the form of NaC$_8$ as shown in Fig.~\ref{fig1}(a). If more Na atoms (e.g. Na$_2$C$_8$) were intercalated into T-carbon, the structure would also  be unstable.

\begin{figure}[!htbp]
  \centering
  \includegraphics[scale=0.75,angle=0]{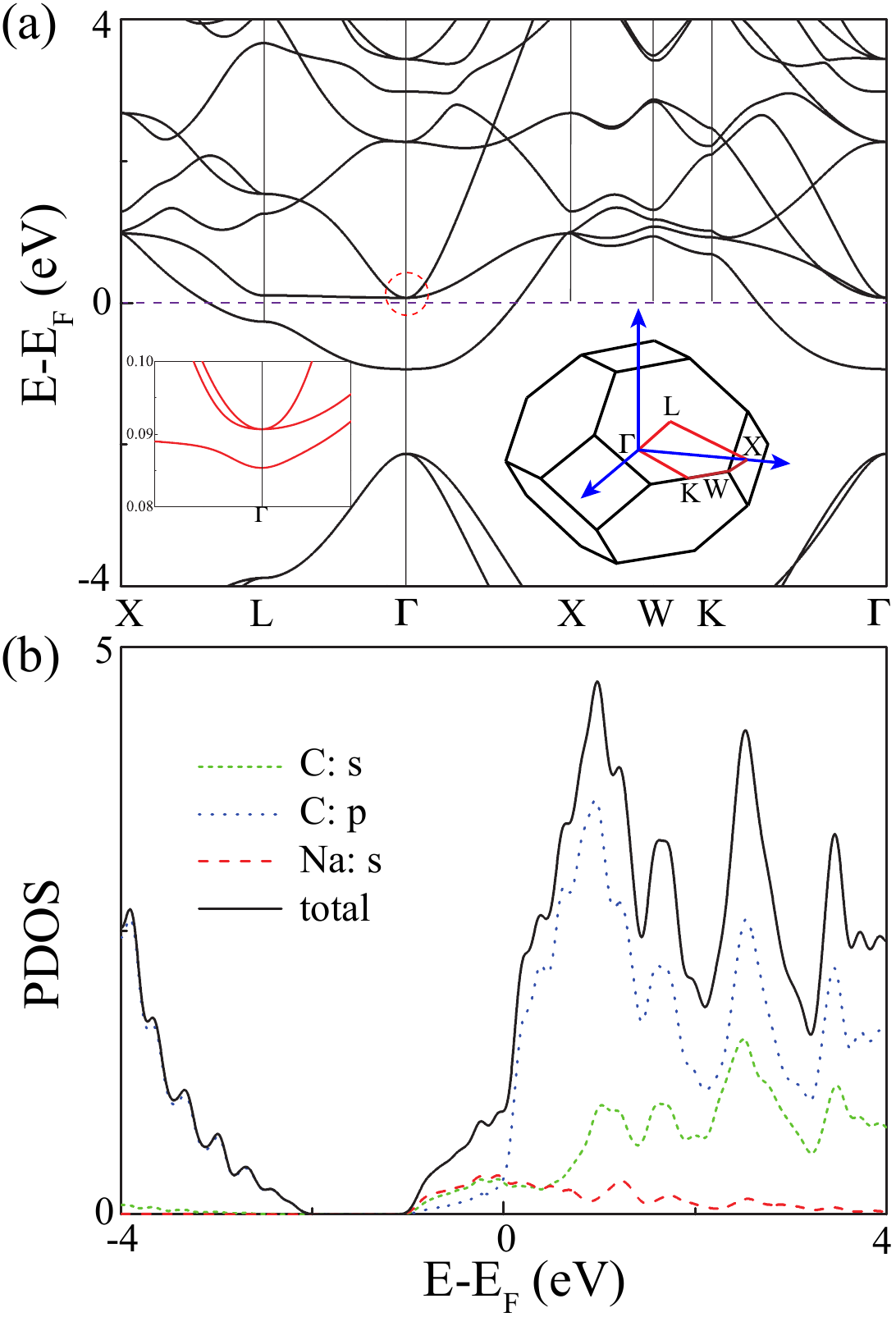}\\
  \caption{(a) The electron band structure and (b) the partial density of states of the Na-doped T-carbon with inclusion of the SOC. The left lower inset in (a) is the enlarged part marked by the red dashed cycle in band structures.}\label{fig3}
\end{figure}

\section{Electronic properties}

The electronic structures and density of states (DOS) of Na-doped T-carbon are given in Fig.~\ref{fig3}. It can be observed that the Na-doped T-carbon is metallic with a dispersive band across the Fermi level. The bands around the Fermi energy are dominated by electrons in $s$ and $p$ orbitals of C atoms and $s$ orbital of Na atom. Without inclusion of spin-orbit coupling (SOC), there is a triple degenerate point at $\Gamma$ point near Fermi level. With including the SOC, the triple degenerate point at $\Gamma$ point turns into four-fold degenerate point, because each band is doubly degenerate that is cased by the symmetries. Moreover, around the degenerate point, the bands have quadratic dispersion along all directions. Such a point is qualified as a quadratic contact point~\cite{Zhu2018}, which can be transformed into a variety topological phases, and has unconventional features in the Landau spectrum under a strong magnetic field~\cite{Zhu2018}. The Fermi surface of Na-doped T-carbon was plotted in Fig.~\ref{fig1}(b). It is a corner-truncated tetrakaidecahedron with an ``electron-type'' structure. The Fermi surface possesses the same symmetry as the crystalline structure of Na-doped T-carbon.

\begin{figure}[!htbp]
  \centering
  \includegraphics[scale=0.75,angle=0]{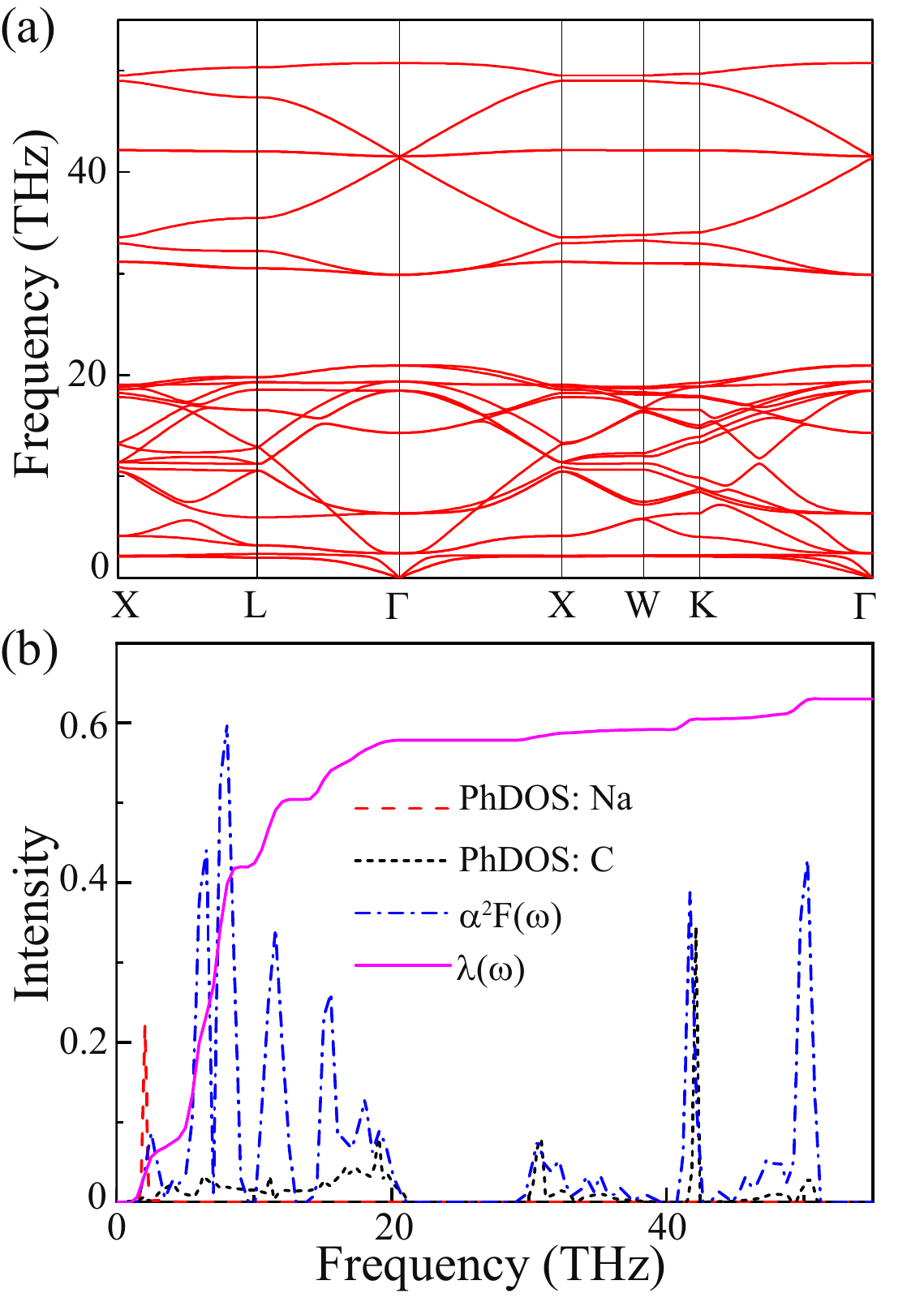}\\
  \caption{(a) The phonon spectra and (b) the phonon density of states (PhDOS) (divided by 40), Eliashberg spectral function $\alpha^2F(\omega)$, and the cumulative frequency-dependent of EPC $\lambda(\omega)$ of Na-doped T-carbon at ambient pressure.}\label{fig4}
\end{figure}

\section{Superconductivity}

As the Na-doped T-carbon is a metal with an electronic Fermi surface mainly contributed by electrons in $s$ and $p$ orbitals, superconductivity may be induced by phonon-mediated electron pairing. We now focus on the phonon properties and the electro-phonon coupling (EPC) of the Na-doped T-carbon. Figure~\ref{fig4}(a) shows the phonon spectra along high-symmetry path of X-L-$\Gamma$-X-W-K-$\Gamma$ for the Na-doped T-carbon. No imaginary frequency mode of phonons is found, indicating that the compound is dynamically stable. It is instructive to note that there exists a wide direct band gap (9 THz) of phonons between the frequency 21-30 THz at $\Gamma$ point. From the animation of phonon modes, we find that the main contribution to the low-frequency modes of the acoustic branches below 2 Thz is from the vibrations of C and Na atoms in (111) plane. The vibrations of Na atoms only locate at low-frequency with a peak of phonon density of states (PhDOS) at about 2 THz. According to Migdal-Eliashberg theory,~\cite{Guentherodt1980,Giustino2017} the EPC parameter $\lambda_{qv}$ can be calculated by
\begin{equation}
\lambda_{qv}=\frac{\gamma_{qv}}{\pi h N(E_F)\omega_{qv}^2},
\end{equation}
where $\gamma_{qv}$ is the phonon linewidth, $\omega_{qv}$ is the phonon frequency, and N(E$_F$) is the electronic density of states at the Fermi level. $\gamma_{qv}$ can be estimated by
\begin{equation}
\gamma_{qv}=\frac{2\pi\omega_{qv}}{\Omega_{BZ}}\sum_{k,n,m}|g_{kn,k+qm}^\nu|^2\delta(\epsilon_{kn}-\epsilon_F)\delta(\epsilon_{k+qm}-\epsilon_F),
\end{equation}
where $\Omega_{BZ}$ is the volume of BZ, $\epsilon_{kn}$ and $\epsilon_{k+qm}$ denote the Kohn-Sham energy, and $g_{kn,k+qm}^\nu$ represents the EPC matrix element, which describes the probability amplitude for the scattering of an electron with a transfer of crystal momentum $q$ and can be obtained self-consistently by the linear response theory~\cite{Allen1975}. The Eliashberg electron-phonon spectral function $\alpha^2F(\omega)$, and the
cumulative frequency-dependent EPC $\lambda(\omega)$ can be calculated by
\begin{equation}
\alpha^2F(\omega)=\frac{1}{2\pi {\rm N(E}_F)}\sum_{q\nu}\frac{\gamma_{q\nu}}{\omega_{q\nu}}\delta(\omega-\omega_{q\nu}),
\end{equation}
and
\begin{equation}
\lambda(\omega)=2\int_0^\omega \frac{\alpha^2F(\omega)}{\omega}d\omega,
\end{equation}
respectively.

The phonon spectra as well as the phonon density of states (PhDOS), $\alpha^2F(\omega)$, and $\lambda(\omega)$ are presented in Fig.~\ref{fig4}(b). From the PhDOS, one can observe a peak from Na atoms at the low frequency. It is noted that the low-frequency phonons (below 10 THz) account for 67\% of the total EPC ($\lambda$ = $\lambda(\infty)$ = 0.63), while the phonons below 20 THz contribute 92\% to the total EPC. The main parts of the PhDOS and $\alpha^2F(\omega)$ are also distributed in this region, while the phonons in high-frequency region contribute little to the EPC strength.

Utilizing our calculated $\alpha^2F(\omega)$ and $\lambda(\omega)$ together with a typical value of the effective screened Coulomb repulsion constant $\mu^*$ = 0.1, we calculate the logarithmic average frequency $\omega_{log}$ by
\begin{equation}
\omega_{log}={\rm exp}[{\frac{2}{\lambda}\int_0^\infty\frac{d\omega}{\omega}\alpha^2F(\omega)log\omega}].
\end{equation}
The superconducting transition temperature Tc can be obtained by
\begin{equation}
{\rm Tc}=\frac{\omega_{log}}{1.2}{\rm exp}[{-\frac{1.04(1+\lambda)}{\lambda-\mu^*(1+0.62\lambda)}}].
\end{equation}
The relative parameters of N(E$_F$), $\omega_{log}$, $\lambda$, and Tc for Na-doped T-carbon are listed in Table~\ref{tab:1}. Tc of 10.9 K was obtained for Na-doped T-carbon at ambient pressure,  indicating that it may be a novel carbon-based superconductor. In analogy to alkali-doped fullerides~\cite{Nomura2016}, the Na-doped T-carbon could be a s-wave BCS superconductor owing to its highly symmetric electronic type Fermi surface.

\section{Thermodynamic properties in normal state}

\begin{figure}[!htbp]
  \centering
  \includegraphics[scale=0.5,angle=0]{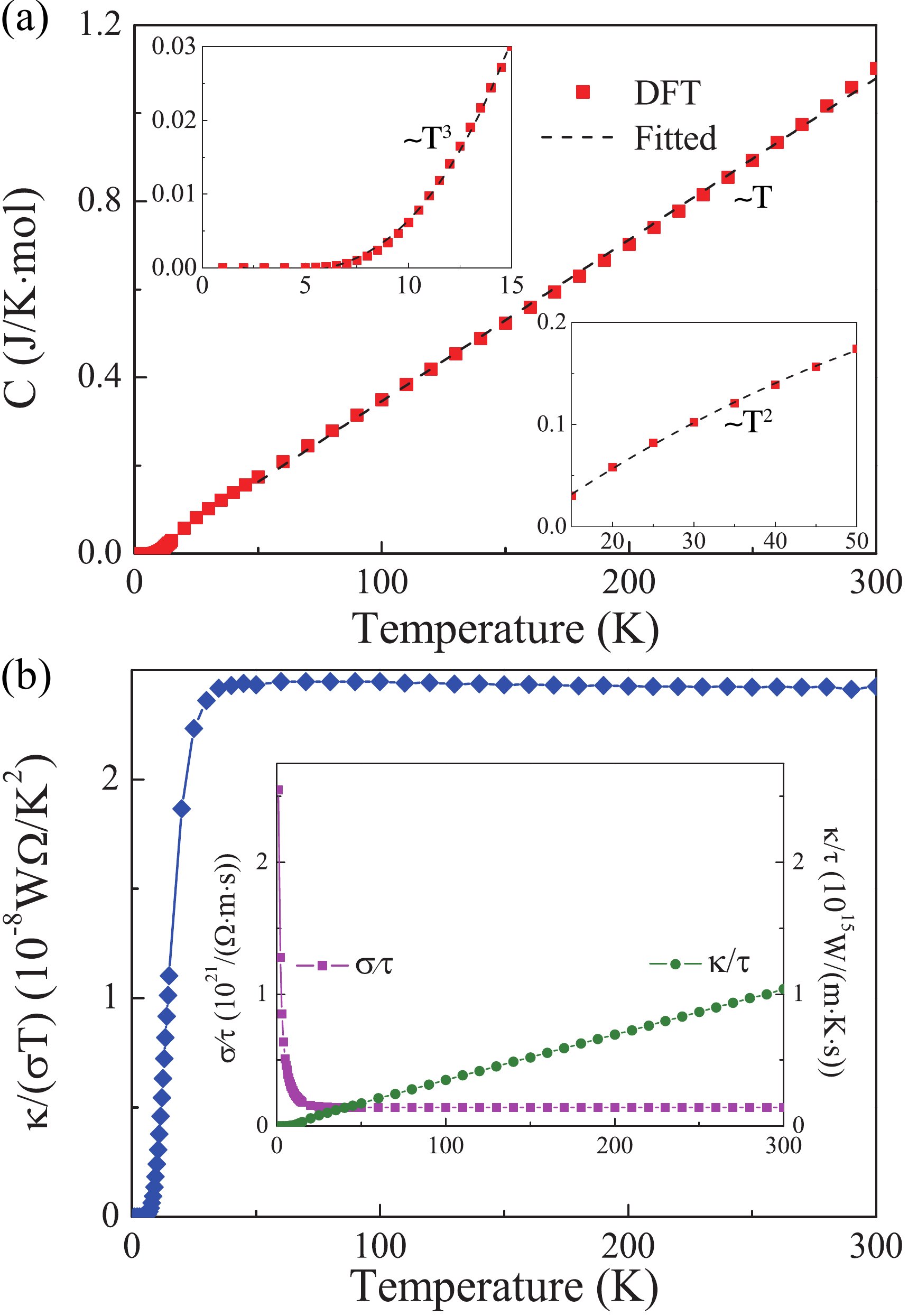}\\
  \caption{(a) Temperature dependence of electronic specific heat C in the normal state of Na-doped T-carbon superconductor. The upper and lower insets are the enlarged parts of the specific heat at temperature T $<$ 15 K and 15 K $<$ T $<$ 50 K, respectively.
The dash lines are fitting curves at different temperature regions. (b) Temperature dependent Lorenz number $L$ [$L=\kappa$/($\sigma$T)]. The inset exhibits the temperature dependent electrical ($\sigma$) and thermal ($\kappa$) conductivities over the relaxation time ($\tau$). }\label{fig5}
\end{figure}

The temperature dependence of the electronic specific heat of the Na-doped T-carbon superconductor in normal state is calculated, as presented in Fig.~\ref{fig5}(a). It is seen that the specific heat C(T) has distinct behaviors. At low temperature (T$<$15 K), C(T)$\sim$T$^3$ (upper inset), at 15 K$<$T$<$50 K, C(T)$\sim$T$^2$ (lower inset), and at T$>$50K, C(T)$\sim$T. It is clear that below 50 K, the normal state shows a non-Fermi liquid behavior.

To further verify this observation, we also studied the temperature dependence of the Lorenz number defined by $L = \kappa/(\sigma T)$, as given in Fig.~\ref{fig5}(b), where the electrical conductivity $\sigma$ over relaxation time $\tau$ and the thermal conductivity $\kappa$ over $\tau$ are calculated, as presented in the inset. It can be observed that at low temperature (T$<$50 K), $\kappa/(\sigma T)$ is not a constant, showing that it violates dramatically the Wiedemann-Franz law, and at high temperature it shows almost a constant, indicating a Fermi liquid behavior.

Therefore, the normal state of Na-doped T-carbon superconductor reveals a non-Fermi liquid behavior at temperature less than 50 K, implying that the interactions between electrons in the normal state play essential roles at low temperature. In addition, one may notice that the $T^3$-behavior of specific heat at low temperature resembles the characteristic nature of bosonic quasiparticles, suggesting that such a non-Fermi liquid normal state might be an anomalous metal.

\section{Pressure effect in superconducting state}

\begin{figure}[!htbp]
  \centering
  \includegraphics[scale=0.7,angle=0]{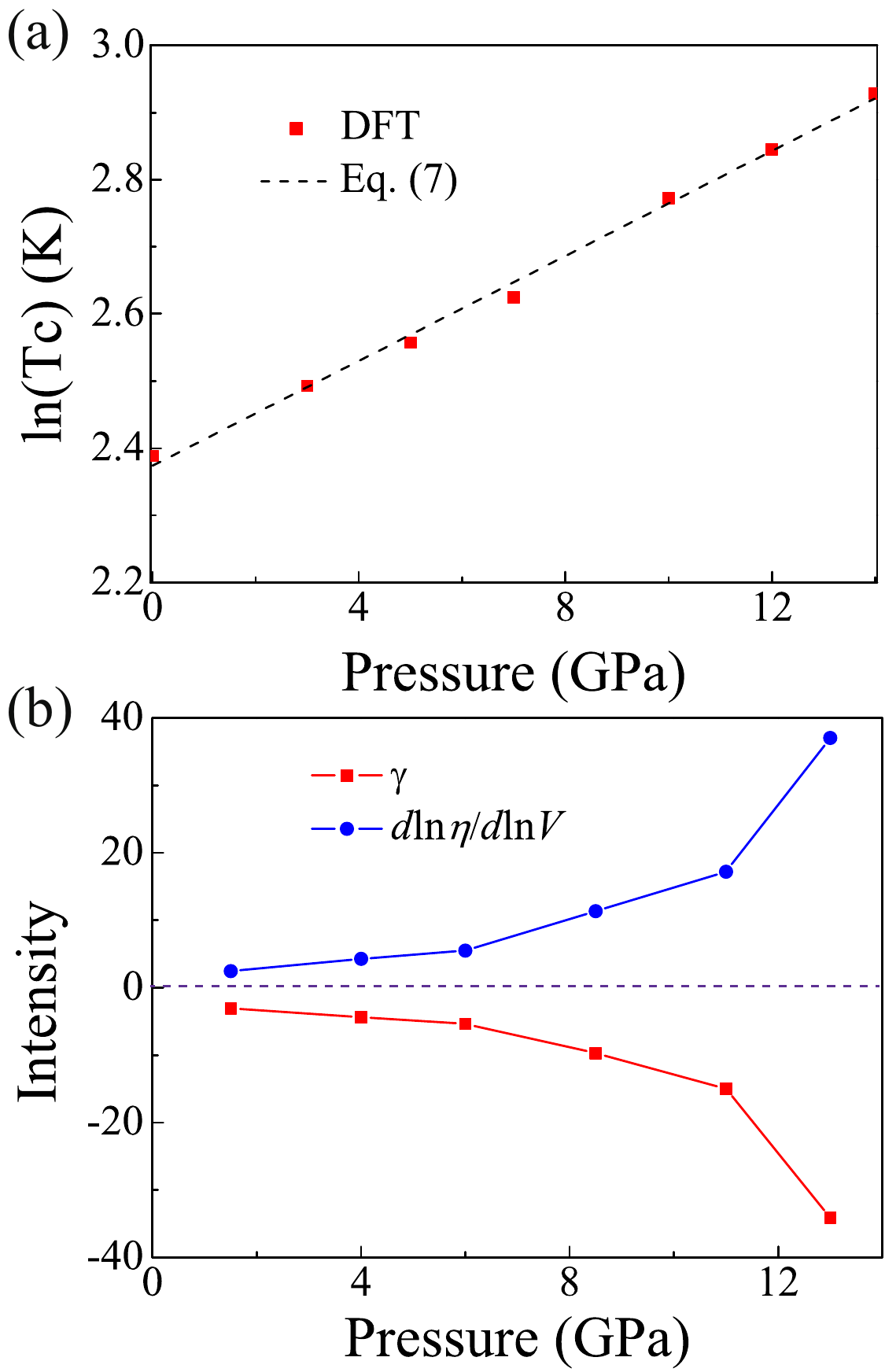}\\
  \caption{Pressure dependence of (a) logarithmic superconducting transition temperature, (b) Hopfield ($d$ln$\eta$/$d$ln$V$) and Gr\"{u}neisen ($\gamma$) parameters in Eq.~(\ref{eq-Tc}) for the Na-doped T-carbon superconductor.}\label{fig6}
\end{figure}

To study the effect of pressure on superconducting transition temperature Tc for Na-doped T-carbon superconductor, we calculated several low-pressure cases (below 14 GPa), where the results are listed in Table~\ref{tab:1}. As can be seen, with the increase of pressure, the cumulative EPC $\lambda$ increases while $\alpha^2F(\omega)$ decreases, and both together lead to the increase of Tc. The pressure dependent of Tc(P) is presented in Fig.~\ref{fig6}(a).

\begin{table}[htbp]
  \caption{The pressure dependent superconductive parameters of N(E$_F$) (in unit of states/spin/eV/cell), volume (\AA$^3$), $\omega_{log}$ (in K), $\lambda$, and Tc (in K) for Na-doped T-carbon.}
  \label{tab:1}
  \begin{tabular}{cccccccc}
		\hline
		P(GPa)   &N(E$_F$) &V(\AA$^3$)  &$\omega_{log}$(K)   &$\lambda$ &Tc(K) \\
		\hline  \hline
		0    &0.85  &108.85   &413.2     &0.63     &10.9          \\
		3    &0.83  &106.81   &389.8     &0.67     &12.1        \\
        5    &0.82  &105.54   &369.8     &0.70     &12.9       \\
		7    &0.82  &104.35   &348.0     &0.74     &13.8       \\
        10   &0.80  &102.67   &297.2     &0.86     &16.0          \\
        12   &0.79  &101.61   &254.4     &0.98     &17.2             \\
        14   &0.79  &100.60   &181.0     &1.36     &18.7           \\
		\hline
	\end{tabular}
\end{table}

Tc(P) obeys the following relation~\cite{Schillinga}
\begin{equation}\label{eq-Tc}
\frac{d \rm ln T_c}{d {\rm ln} V}=-B\frac{d \rm ln T_c}{d P}=-\gamma+\Delta \{\frac{d {\rm ln} \eta}{d {\rm ln} V}+2\gamma\},
\end{equation}
where B is the bulk modulus ($\sim$ 178 GPa), $\gamma$ $\equiv$ -$d$ln$\langle \omega \rangle$/$d$ln$V$ is the Gr\"{u}neisen parameter, $\eta$ $\equiv$ N(E$_f$)$\langle I^2\rangle$ is the Hopfield parameter~\cite{Hopfield1971} with $\langle I^2\rangle$ the square of the electron-phonon matrix element averaged over the Fermi surface, and $\Delta$ $\equiv$ 1.04$\lambda$(1+0.38$\mu^*$)[$\lambda$-$\mu^*$(1+0.62$\lambda$)]$^{-2}$. Due to the first term on the right hand side of Eq.~(\ref{eq-Tc}) smaller than the second~\cite{Chen2002}, the sign of $d$Tc/$dP$ is determined by the relative magnitude of the two terms in curly brackets. The Gr\"{u}neisen parameter ($\gamma$) can  be directly obtained from Table~\ref{tab:1}, and the Hopfield term can be determined by Eq.~(\ref{eq-Tc}). These results are shown in Fig.~\ref{fig6}. It can be seen that our calculated results are perfectly in agreement with Eq.~(\ref{eq-Tc}).
From Fig.~\ref{fig6}(a), we can obtain $d$lnTc/$d$ln$V$=-6.94 based on $B$ and $d$lnTc/$dP$(=0.039), where the Gr\"{u}neisen parameter $\gamma<$0 and Hopfield term $d$ln$\eta$/$d$ln$V$$>$0 [Fig.~\ref{fig6}(b)], the latter originates from the slightly decrease of electronic density of states N(E$_f$) under pressure as given in Table~\ref{tab:1}. Therefore, we may see that the calculated increase in Tc with pressure for Na-doped T-carbon superconductor results from an enhancement of the electron-phonon coupling $\lambda$ due to the shift of the phonon spectrum to lower frequencies.

It is worthy to mention that the pressure dependent behavior of Tc in Na-doped T-carbon is uncovered in contrast to the doped fullerenes A$_3$C$_{60}$ (A=K, Rb, Cs), where the transition temperature decreases under pressure~\cite{Tanigaki1991,Palstra1995,ruoff1997recent,Diederichs1996,Diederichs1997}. The reason is that a rapid decrease of Tc under pressure in A$_3$C$_{60}$ is owing to the sharp decrease of electronic density of states N(E$_f$) under pressure, because of a rapid increase of the width of the conduction band when C$_{60}$ molecules are pressed together.

\section{Summary}

In this work, we show that the Na-doped T-carbon is a phonon-mediated superconductor in accordance with the BCS mechanism. The superconducting transition temperature Tc is estimated to be 11 K under ambient pressure, and can be reached to 19 K at the pressure of 14 GPa. The increase of Tc under pressure was unveiled from the enhancement of the electron-phonon coupling owing to the shift of the phonon spectrum to lower frequencies. It is observed that the increased behavior of superconducting transition temperature of Na-doped T-carbon under pressure is in contrast to the A$_3$C$_{60}$ superconductors, where Tc decreases under pressure because of the sharp decrease of electronic density of states N(E$_f$) under pressure. The low-temperature behaviors of specific heat as well as electrical and thermal conductivities indicate that the normal state of Na-doped T-carbon superconductor is a non-Fermi liquid below 50 K, implying that the interactions between electrons may play roles in the normal state at low temperature. Our results would spur great interest both experimentally and theoretically to explore novel carbon-based superconductors and deepen our understandings on novel properties of carbon.

\section*{Appendix A. Eliashberg spectral functions $\alpha^2F(\omega)$ under pressure}

\begin{figure}[!!!htbp]
  \centering
  \includegraphics[scale=0.78,angle=0]{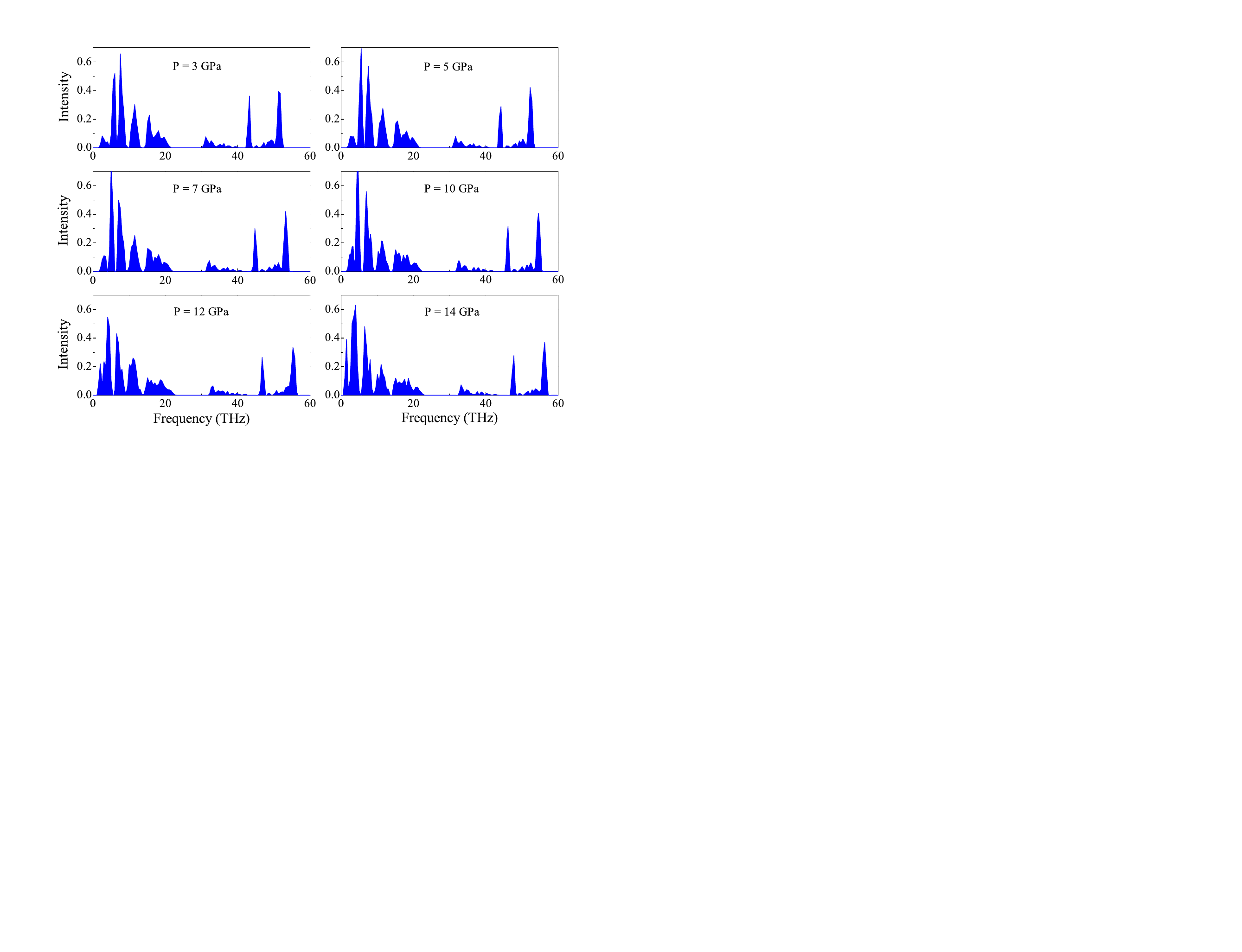}\\
  \caption{The Eliashberg spectral functions $\alpha^2F(\omega)$ under various pressure.}\label{fig7}
\end{figure}

The Eliashberg spectral functions $\alpha^2F(\omega)$ under various pressure are given in Fig.~\ref{fig7} for further information. It is obvious that the peaks of $\alpha^2F(\omega)$ move to both sides of frequency with increasing pressure, i.e the low-frequency (below 20 THz) phonons move to the lower frequency region and the high-frequency (above 50 THz) phonons move to the higher frequency region. As seen from Fig.~\ref{fig4}(b), the total EPC $\lambda$ is mainly from the low-frequency phonons (below 20 THz), thus the shift of the
$\alpha^2F(\omega)$ to lower frequencies may lead to an increase of $\lambda$, as presented in Table~\ref{tab:1}.

\acknowledgements

This work is supported in part by the National Key R\&D Program of China (Grant No. 2018YFA0305800), the Strategic Priority Research Program of the Chinese Academy of Sciences (Grant No. XDB28000000), the National Natural Science Foundation of China (Grant No.11834014), and Beijing Municipal Science and Technology Commission (Grant No. Z118100004218001). B.G. is also supported by the National Natural Science Foundation of China (Grant No. Y81Z01A1A9), the Chinese Academy of Sciences (Grant No. Y929013EA2), the University of Chinese Academy of Sciences (Grant No. 110200M208), and the Beijing Natural Science Foundation (Grant No. Z190011).


\begin{thebibliography}{48}%
\makeatletter
\providecommand \@ifxundefined [1]{%
 \@ifx{#1\undefined}
}%
\providecommand \@ifnum [1]{%
 \ifnum #1\expandafter \@firstoftwo
 \else \expandafter \@secondoftwo
 \fi
}%
\providecommand \@ifx [1]{%
 \ifx #1\expandafter \@firstoftwo
 \else \expandafter \@secondoftwo
 \fi
}%
\providecommand \natexlab [1]{#1}%
\providecommand \enquote  [1]{``#1''}%
\providecommand \bibnamefont  [1]{#1}%
\providecommand \bibfnamefont [1]{#1}%
\providecommand \citenamefont [1]{#1}%
\providecommand \href@noop [0]{\@secondoftwo}%
\providecommand \href [0]{\begingroup \@sanitize@url \@href}%
\providecommand \@href[1]{\@@startlink{#1}\@@href}%
\providecommand \@@href[1]{\endgroup#1\@@endlink}%
\providecommand \@sanitize@url [0]{\catcode `\\12\catcode `\$12\catcode
  `\&12\catcode `\#12\catcode `\^12\catcode `\_12\catcode `\%12\relax}%
\providecommand \@@startlink[1]{}%
\providecommand \@@endlink[0]{}%
\providecommand \url  [0]{\begingroup\@sanitize@url \@url }%
\providecommand \@url [1]{\endgroup\@href {#1}{\urlprefix }}%
\providecommand \urlprefix  [0]{URL }%
\providecommand \Eprint [0]{\href }%
\providecommand \doibase [0]{https://doi.org/}%
\providecommand \selectlanguage [0]{\@gobble}%
\providecommand \bibinfo  [0]{\@secondoftwo}%
\providecommand \bibfield  [0]{\@secondoftwo}%
\providecommand \translation [1]{[#1]}%
\providecommand \BibitemOpen [0]{}%
\providecommand \bibitemStop [0]{}%
\providecommand \bibitemNoStop [0]{.\EOS\space}%
\providecommand \EOS [0]{\spacefactor3000\relax}%
\providecommand \BibitemShut  [1]{\csname bibitem#1\endcsname}%
\let\auto@bib@innerbib\@empty
\bibitem [{\citenamefont {Lauginie}\ \emph {et~al.}(1980)\citenamefont
  {Lauginie}, \citenamefont {Estrade}, \citenamefont {Conard}, \citenamefont
  {Gu{\'{e}}rard}, \citenamefont {Lagrange},\ and\ \citenamefont
  {Makrini}}]{Lauginie1980}%
  \BibitemOpen
  \bibfield  {author} {\bibinfo {author} {\bibfnamefont {P.}~\bibnamefont
  {Lauginie}}, \bibinfo {author} {\bibfnamefont {H.}~\bibnamefont {Estrade}},
  \bibinfo {author} {\bibfnamefont {J.}~\bibnamefont {Conard}}, \bibinfo
  {author} {\bibfnamefont {D.}~\bibnamefont {Gu{\'{e}}rard}}, \bibinfo {author}
  {\bibfnamefont {P.}~\bibnamefont {Lagrange}},\ and\ \bibinfo {author}
  {\bibfnamefont {M.~E.}\ \bibnamefont {Makrini}},\ }\bibfield  {title}
  {\bibinfo {title} {Graphite lamellar compounds {EPR} studies},\ }\href
  {https://doi.org/10.1016/0378-4363(80)90288-0} {\bibfield  {journal}
  {\bibinfo  {journal} {Physica B}\ }\textbf {\bibinfo {volume} {99}},\
  \bibinfo {pages} {514} (\bibinfo {year} {1980})}\BibitemShut {NoStop}%
\bibitem [{\citenamefont {Ekimov}\ \emph {et~al.}(2004)\citenamefont {Ekimov},
  \citenamefont {Sidorov}, \citenamefont {Bauer}, \citenamefont
  {Mel{\textquotesingle}nik}, \citenamefont {Curro}, \citenamefont {Thompson},\
  and\ \citenamefont {Stishov}}]{Ekimov2004}%
  \BibitemOpen
  \bibfield  {author} {\bibinfo {author} {\bibfnamefont {E.~A.}\ \bibnamefont
  {Ekimov}}, \bibinfo {author} {\bibfnamefont {V.~A.}\ \bibnamefont {Sidorov}},
  \bibinfo {author} {\bibfnamefont {E.~D.}\ \bibnamefont {Bauer}}, \bibinfo
  {author} {\bibfnamefont {N.~N.}\ \bibnamefont {Mel{\textquotesingle}nik}},
  \bibinfo {author} {\bibfnamefont {N.~J.}\ \bibnamefont {Curro}}, \bibinfo
  {author} {\bibfnamefont {J.~D.}\ \bibnamefont {Thompson}},\ and\ \bibinfo
  {author} {\bibfnamefont {S.~M.}\ \bibnamefont {Stishov}},\ }\bibfield
  {title} {\bibinfo {title} {Superconductivity in diamond},\ }\href
  {https://doi.org/10.1038/nature02449} {\bibfield  {journal} {\bibinfo
  {journal} {Nature}\ }\textbf {\bibinfo {volume} {428}},\ \bibinfo {pages}
  {542} (\bibinfo {year} {2004})}\BibitemShut {NoStop}%
\bibitem [{\citenamefont {Kroto}\ \emph {et~al.}(1985)\citenamefont {Kroto},
  \citenamefont {Heath}, \citenamefont {O{\textquotesingle}Brien},
  \citenamefont {Curl},\ and\ \citenamefont {Smalley}}]{Kroto1985}%
  \BibitemOpen
  \bibfield  {author} {\bibinfo {author} {\bibfnamefont {H.~W.}\ \bibnamefont
  {Kroto}}, \bibinfo {author} {\bibfnamefont {J.~R.}\ \bibnamefont {Heath}},
  \bibinfo {author} {\bibfnamefont {S.~C.}\ \bibnamefont
  {O{\textquotesingle}Brien}}, \bibinfo {author} {\bibfnamefont {R.~F.}\
  \bibnamefont {Curl}},\ and\ \bibinfo {author} {\bibfnamefont {R.~E.}\
  \bibnamefont {Smalley}},\ }\bibfield  {title} {\bibinfo {title} {C60:
  Buckminsterfullerene},\ }\href {https://doi.org/10.1038/318162a0} {\bibfield
  {journal} {\bibinfo  {journal} {Nature}\ }\textbf {\bibinfo {volume} {318}},\
  \bibinfo {pages} {162} (\bibinfo {year} {1985})}\BibitemShut {NoStop}%
\bibitem [{\citenamefont {Krätschmer}\ \emph {et~al.}(1990)\citenamefont
  {Krätschmer}, \citenamefont {Lamb}, \citenamefont {Fostiropoulos},\ and\
  \citenamefont {Huffman}}]{Kraetschmer1990}%
  \BibitemOpen
  \bibfield  {author} {\bibinfo {author} {\bibfnamefont {W.}~\bibnamefont
  {Krätschmer}}, \bibinfo {author} {\bibfnamefont {L.~D.}\ \bibnamefont
  {Lamb}}, \bibinfo {author} {\bibfnamefont {K.}~\bibnamefont
  {Fostiropoulos}},\ and\ \bibinfo {author} {\bibfnamefont {D.~R.}\
  \bibnamefont {Huffman}},\ }\bibfield  {title} {\bibinfo {title} {Solid c60: a
  new form of carbon},\ }\href {https://doi.org/10.1038/347354a0} {\bibfield
  {journal} {\bibinfo  {journal} {Nature}\ }\textbf {\bibinfo {volume} {347}},\
  \bibinfo {pages} {354} (\bibinfo {year} {1990})}\BibitemShut {NoStop}%
\bibitem [{\citenamefont {Haddon}\ \emph {et~al.}(1991)\citenamefont {Haddon},
  \citenamefont {Hebard}, \citenamefont {Rosseinsky}, \citenamefont {Murphy},
  \citenamefont {Duclos}, \citenamefont {Lyons}, \citenamefont {Miller},
  \citenamefont {Rosamilia}, \citenamefont {Fleming}, \citenamefont {Kortan},
  \citenamefont {Glarum}, \citenamefont {Makhija}, \citenamefont {Muller},
  \citenamefont {Eick}, \citenamefont {Zahurak}, \citenamefont {Tycko},
  \citenamefont {Dabbagh},\ and\ \citenamefont {Thiel}}]{Haddon1991}%
  \BibitemOpen
  \bibfield  {author} {\bibinfo {author} {\bibfnamefont {R.~C.}\ \bibnamefont
  {Haddon}}, \bibinfo {author} {\bibfnamefont {A.~F.}\ \bibnamefont {Hebard}},
  \bibinfo {author} {\bibfnamefont {M.~J.}\ \bibnamefont {Rosseinsky}},
  \bibinfo {author} {\bibfnamefont {D.~W.}\ \bibnamefont {Murphy}}, \bibinfo
  {author} {\bibfnamefont {S.~J.}\ \bibnamefont {Duclos}}, \bibinfo {author}
  {\bibfnamefont {K.~B.}\ \bibnamefont {Lyons}}, \bibinfo {author}
  {\bibfnamefont {B.}~\bibnamefont {Miller}}, \bibinfo {author} {\bibfnamefont
  {J.~M.}\ \bibnamefont {Rosamilia}}, \bibinfo {author} {\bibfnamefont {R.~M.}\
  \bibnamefont {Fleming}}, \bibinfo {author} {\bibfnamefont {A.~R.}\
  \bibnamefont {Kortan}}, \bibinfo {author} {\bibfnamefont {S.~H.}\
  \bibnamefont {Glarum}}, \bibinfo {author} {\bibfnamefont {A.~V.}\
  \bibnamefont {Makhija}}, \bibinfo {author} {\bibfnamefont {A.~J.}\
  \bibnamefont {Muller}}, \bibinfo {author} {\bibfnamefont {R.~H.}\
  \bibnamefont {Eick}}, \bibinfo {author} {\bibfnamefont {S.~M.}\ \bibnamefont
  {Zahurak}}, \bibinfo {author} {\bibfnamefont {R.}~\bibnamefont {Tycko}},
  \bibinfo {author} {\bibfnamefont {G.}~\bibnamefont {Dabbagh}},\ and\ \bibinfo
  {author} {\bibfnamefont {F.~A.}\ \bibnamefont {Thiel}},\ }\bibfield  {title}
  {\bibinfo {title} {Conducting films of c60 and c70 by alkali-metal doping},\
  }\href {https://doi.org/10.1038/350320a0} {\bibfield  {journal} {\bibinfo
  {journal} {Nature}\ }\textbf {\bibinfo {volume} {350}},\ \bibinfo {pages}
  {320} (\bibinfo {year} {1991})}\BibitemShut {NoStop}%
\bibitem [{\citenamefont {Hebard}\ \emph {et~al.}(1991)\citenamefont {Hebard},
  \citenamefont {Rosseinsky}, \citenamefont {Haddon}, \citenamefont {Murphy},
  \citenamefont {Glarum}, \citenamefont {Palstra}, \citenamefont {Ramirez},\
  and\ \citenamefont {Kortan}}]{Hebard1991}%
  \BibitemOpen
  \bibfield  {author} {\bibinfo {author} {\bibfnamefont {A.~F.}\ \bibnamefont
  {Hebard}}, \bibinfo {author} {\bibfnamefont {M.~J.}\ \bibnamefont
  {Rosseinsky}}, \bibinfo {author} {\bibfnamefont {R.~C.}\ \bibnamefont
  {Haddon}}, \bibinfo {author} {\bibfnamefont {D.~W.}\ \bibnamefont {Murphy}},
  \bibinfo {author} {\bibfnamefont {S.~H.}\ \bibnamefont {Glarum}}, \bibinfo
  {author} {\bibfnamefont {T.~T.~M.}\ \bibnamefont {Palstra}}, \bibinfo
  {author} {\bibfnamefont {A.~P.}\ \bibnamefont {Ramirez}},\ and\ \bibinfo
  {author} {\bibfnamefont {A.~R.}\ \bibnamefont {Kortan}},\ }\bibfield  {title}
  {\bibinfo {title} {Superconductivity at 18 k in potassium-doped c60},\ }\href
  {https://doi.org/10.1038/350600a0} {\bibfield  {journal} {\bibinfo  {journal}
  {Nature}\ }\textbf {\bibinfo {volume} {350}},\ \bibinfo {pages} {600}
  (\bibinfo {year} {1991})}\BibitemShut {NoStop}%
\bibitem [{\citenamefont {Hebard}(1992)}]{Hebard1992}%
  \BibitemOpen
  \bibfield  {author} {\bibinfo {author} {\bibfnamefont {A.~F.}\ \bibnamefont
  {Hebard}},\ }\bibfield  {title} {\bibinfo {title} {Superconductivity in doped
  fullerenes},\ }\href@noop {} {\bibfield  {journal} {\bibinfo  {journal}
  {Phys. Today}\ }\textbf {\bibinfo {volume} {11}},\ \bibinfo {pages} {26}
  (\bibinfo {year} {1992})}\BibitemShut {NoStop}%
\bibitem [{\citenamefont {Kortan}\ \emph
  {et~al.}(1992{\natexlab{a}})\citenamefont {Kortan}, \citenamefont {Kopylov},
  \citenamefont {Glarum}, \citenamefont {Gyorgy}, \citenamefont {Ramirez},
  \citenamefont {Fleming}, \citenamefont {Thiel},\ and\ \citenamefont
  {Haddon}}]{Kortan1992}%
  \BibitemOpen
  \bibfield  {author} {\bibinfo {author} {\bibfnamefont {A.~R.}\ \bibnamefont
  {Kortan}}, \bibinfo {author} {\bibfnamefont {N.}~\bibnamefont {Kopylov}},
  \bibinfo {author} {\bibfnamefont {S.}~\bibnamefont {Glarum}}, \bibinfo
  {author} {\bibfnamefont {E.~M.}\ \bibnamefont {Gyorgy}}, \bibinfo {author}
  {\bibfnamefont {A.~P.}\ \bibnamefont {Ramirez}}, \bibinfo {author}
  {\bibfnamefont {R.~M.}\ \bibnamefont {Fleming}}, \bibinfo {author}
  {\bibfnamefont {F.~A.}\ \bibnamefont {Thiel}},\ and\ \bibinfo {author}
  {\bibfnamefont {R.~C.}\ \bibnamefont {Haddon}},\ }\bibfield  {title}
  {\bibinfo {title} {Superconductivity at 8.4 k in calcium-doped c60},\ }\href
  {https://doi.org/10.1038/355529a0} {\bibfield  {journal} {\bibinfo  {journal}
  {Nature}\ }\textbf {\bibinfo {volume} {355}},\ \bibinfo {pages} {529}
  (\bibinfo {year} {1992}{\natexlab{a}})}\BibitemShut {NoStop}%
\bibitem [{\citenamefont {Kortan}\ \emph
  {et~al.}(1992{\natexlab{b}})\citenamefont {Kortan}, \citenamefont {Kopylov},
  \citenamefont {Glarum}, \citenamefont {Gyorgy}, \citenamefont {Ramirez},
  \citenamefont {Fleming}, \citenamefont {Zhou}, \citenamefont {Thiel},
  \citenamefont {Trevor},\ and\ \citenamefont {Haddon}}]{Kortan1992a}%
  \BibitemOpen
  \bibfield  {author} {\bibinfo {author} {\bibfnamefont {A.~R.}\ \bibnamefont
  {Kortan}}, \bibinfo {author} {\bibfnamefont {N.}~\bibnamefont {Kopylov}},
  \bibinfo {author} {\bibfnamefont {S.}~\bibnamefont {Glarum}}, \bibinfo
  {author} {\bibfnamefont {E.~M.}\ \bibnamefont {Gyorgy}}, \bibinfo {author}
  {\bibfnamefont {A.~P.}\ \bibnamefont {Ramirez}}, \bibinfo {author}
  {\bibfnamefont {R.~M.}\ \bibnamefont {Fleming}}, \bibinfo {author}
  {\bibfnamefont {O.}~\bibnamefont {Zhou}}, \bibinfo {author} {\bibfnamefont
  {F.~A.}\ \bibnamefont {Thiel}}, \bibinfo {author} {\bibfnamefont {P.~L.}\
  \bibnamefont {Trevor}},\ and\ \bibinfo {author} {\bibfnamefont {R.~C.}\
  \bibnamefont {Haddon}},\ }\bibfield  {title} {\bibinfo {title}
  {Superconductivity in barium fulleride},\ }\href
  {https://doi.org/10.1038/360566a0} {\bibfield  {journal} {\bibinfo  {journal}
  {Nature}\ }\textbf {\bibinfo {volume} {360}},\ \bibinfo {pages} {566}
  (\bibinfo {year} {1992}{\natexlab{b}})}\BibitemShut {NoStop}%
\bibitem [{\citenamefont {Özda{\c{s}}}\ \emph {et~al.}(1995)\citenamefont
  {Özda{\c{s}}}, \citenamefont {Kortan}, \citenamefont {Kopylov},
  \citenamefont {Ramirez}, \citenamefont {Siegrist}, \citenamefont {Rabe},
  \citenamefont {Bair}, \citenamefont {Schuppler},\ and\ \citenamefont
  {Citrin}}]{Oezdas1995}%
  \BibitemOpen
  \bibfield  {author} {\bibinfo {author} {\bibfnamefont {E.}~\bibnamefont
  {Özda{\c{s}}}}, \bibinfo {author} {\bibfnamefont {A.~R.}\ \bibnamefont
  {Kortan}}, \bibinfo {author} {\bibfnamefont {N.}~\bibnamefont {Kopylov}},
  \bibinfo {author} {\bibfnamefont {A.~P.}\ \bibnamefont {Ramirez}}, \bibinfo
  {author} {\bibfnamefont {T.}~\bibnamefont {Siegrist}}, \bibinfo {author}
  {\bibfnamefont {K.~M.}\ \bibnamefont {Rabe}}, \bibinfo {author}
  {\bibfnamefont {H.~E.}\ \bibnamefont {Bair}}, \bibinfo {author}
  {\bibfnamefont {S.}~\bibnamefont {Schuppler}},\ and\ \bibinfo {author}
  {\bibfnamefont {P.~H.}\ \bibnamefont {Citrin}},\ }\bibfield  {title}
  {\bibinfo {title} {Superconductivity and cation-vacancy ordering in the
  rare-earth fulleride yb2.75c60},\ }\href {https://doi.org/10.1038/375126a0}
  {\bibfield  {journal} {\bibinfo  {journal} {Nature}\ }\textbf {\bibinfo
  {volume} {375}},\ \bibinfo {pages} {126} (\bibinfo {year}
  {1995})}\BibitemShut {NoStop}%
\bibitem [{\citenamefont {Yoshikawa}\ \emph {et~al.}(1995)\citenamefont
  {Yoshikawa}, \citenamefont {Kuroshima}, \citenamefont {Hirosawa},
  \citenamefont {Tanigaki},\ and\ \citenamefont {Mizuki}}]{Yoshikawa1995}%
  \BibitemOpen
  \bibfield  {author} {\bibinfo {author} {\bibfnamefont {H.}~\bibnamefont
  {Yoshikawa}}, \bibinfo {author} {\bibfnamefont {S.}~\bibnamefont
  {Kuroshima}}, \bibinfo {author} {\bibfnamefont {I.}~\bibnamefont {Hirosawa}},
  \bibinfo {author} {\bibfnamefont {K.}~\bibnamefont {Tanigaki}},\ and\
  \bibinfo {author} {\bibfnamefont {J.}~\bibnamefont {Mizuki}},\ }\bibfield
  {title} {\bibinfo {title} {Eu fulleride formation studied by photoemission
  spectroscopy},\ }\href {https://doi.org/10.1016/0009-2614(95)00445-a}
  {\bibfield  {journal} {\bibinfo  {journal} {Chem. Phys. Lett.}\ }\textbf
  {\bibinfo {volume} {239}},\ \bibinfo {pages} {103} (\bibinfo {year}
  {1995})}\BibitemShut {NoStop}%
\bibitem [{\citenamefont {Rosseinsky}\ \emph {et~al.}(1991)\citenamefont
  {Rosseinsky}, \citenamefont {Ramirez}, \citenamefont {Glarum}, \citenamefont
  {Murphy}, \citenamefont {Haddon}, \citenamefont {Hebard}, \citenamefont
  {Palstra}, \citenamefont {Kortan}, \citenamefont {Zahurak},\ and\
  \citenamefont {Makhija}}]{Rosseinsky1991}%
  \BibitemOpen
  \bibfield  {author} {\bibinfo {author} {\bibfnamefont {M.~J.}\ \bibnamefont
  {Rosseinsky}}, \bibinfo {author} {\bibfnamefont {A.~P.}\ \bibnamefont
  {Ramirez}}, \bibinfo {author} {\bibfnamefont {S.~H.}\ \bibnamefont {Glarum}},
  \bibinfo {author} {\bibfnamefont {D.~W.}\ \bibnamefont {Murphy}}, \bibinfo
  {author} {\bibfnamefont {R.~C.}\ \bibnamefont {Haddon}}, \bibinfo {author}
  {\bibfnamefont {A.~F.}\ \bibnamefont {Hebard}}, \bibinfo {author}
  {\bibfnamefont {T.~T.~M.}\ \bibnamefont {Palstra}}, \bibinfo {author}
  {\bibfnamefont {A.~R.}\ \bibnamefont {Kortan}}, \bibinfo {author}
  {\bibfnamefont {S.~M.}\ \bibnamefont {Zahurak}},\ and\ \bibinfo {author}
  {\bibfnamefont {A.~V.}\ \bibnamefont {Makhija}},\ }\bibfield  {title}
  {\bibinfo {title} {Superconductivity at 28 k {inRbxC}60},\ }\href
  {https://doi.org/10.1103/physrevlett.66.2830} {\bibfield  {journal} {\bibinfo
   {journal} {Phys. Rev. Lett.}\ }\textbf {\bibinfo {volume} {66}},\ \bibinfo
  {pages} {2830} (\bibinfo {year} {1991})}\BibitemShut {NoStop}%
\bibitem [{\citenamefont {Holczer}\ \emph {et~al.}(1991)\citenamefont
  {Holczer}, \citenamefont {Klein}, \citenamefont {m.~Huang}, \citenamefont
  {Kaner}, \citenamefont {j.~Fu}, \citenamefont {Whetten},\ and\ \citenamefont
  {Diederich}}]{Holczer1991}%
  \BibitemOpen
  \bibfield  {author} {\bibinfo {author} {\bibfnamefont {K.}~\bibnamefont
  {Holczer}}, \bibinfo {author} {\bibfnamefont {O.}~\bibnamefont {Klein}},
  \bibinfo {author} {\bibfnamefont {S.}~\bibnamefont {m.~Huang}}, \bibinfo
  {author} {\bibfnamefont {R.~B.}\ \bibnamefont {Kaner}}, \bibinfo {author}
  {\bibfnamefont {K.}~\bibnamefont {j.~Fu}}, \bibinfo {author} {\bibfnamefont
  {R.~L.}\ \bibnamefont {Whetten}},\ and\ \bibinfo {author} {\bibfnamefont
  {F.}~\bibnamefont {Diederich}},\ }\bibfield  {title} {\bibinfo {title}
  {Alkali-fulleride superconductors: Synthesis, composition, and diamagnetic
  shielding},\ }\href {https://doi.org/10.1126/science.252.5009.1154}
  {\bibfield  {journal} {\bibinfo  {journal} {Science}\ }\textbf {\bibinfo
  {volume} {252}},\ \bibinfo {pages} {1154} (\bibinfo {year}
  {1991})}\BibitemShut {NoStop}%
\bibitem [{\citenamefont {Murphy}\ \emph {et~al.}(1992)\citenamefont {Murphy},
  \citenamefont {Rosseinsky}, \citenamefont {Fleming}, \citenamefont {Tycko},
  \citenamefont {Ramirez}, \citenamefont {Haddon}, \citenamefont {Siegrist},
  \citenamefont {Dabbagh}, \citenamefont {Tully},\ and\ \citenamefont
  {Walstedt}}]{Murphy1992}%
  \BibitemOpen
  \bibfield  {author} {\bibinfo {author} {\bibfnamefont {D.}~\bibnamefont
  {Murphy}}, \bibinfo {author} {\bibfnamefont {M.}~\bibnamefont {Rosseinsky}},
  \bibinfo {author} {\bibfnamefont {R.}~\bibnamefont {Fleming}}, \bibinfo
  {author} {\bibfnamefont {R.}~\bibnamefont {Tycko}}, \bibinfo {author}
  {\bibfnamefont {A.}~\bibnamefont {Ramirez}}, \bibinfo {author} {\bibfnamefont
  {R.}~\bibnamefont {Haddon}}, \bibinfo {author} {\bibfnamefont
  {T.}~\bibnamefont {Siegrist}}, \bibinfo {author} {\bibfnamefont
  {G.}~\bibnamefont {Dabbagh}}, \bibinfo {author} {\bibfnamefont
  {J.}~\bibnamefont {Tully}},\ and\ \bibinfo {author} {\bibfnamefont
  {R.}~\bibnamefont {Walstedt}},\ }\bibfield  {title} {\bibinfo {title}
  {Synthesis and characterization of alkali metal fullerides: {AxC}60},\ }\href
  {https://doi.org/10.1016/0022-3697(92)90230-b} {\bibfield  {journal}
  {\bibinfo  {journal} {J. Phys. Chem. Solids}\ }\textbf {\bibinfo {volume}
  {53}},\ \bibinfo {pages} {1321} (\bibinfo {year} {1992})}\BibitemShut
  {NoStop}%
\bibitem [{\citenamefont {Tanigaki}\ \emph {et~al.}(1991)\citenamefont
  {Tanigaki}, \citenamefont {Ebbesen}, \citenamefont {Saito}, \citenamefont
  {Mizuki}, \citenamefont {Tsai}, \citenamefont {Kubo},\ and\ \citenamefont
  {Kuroshima}}]{Tanigaki1991}%
  \BibitemOpen
  \bibfield  {author} {\bibinfo {author} {\bibfnamefont {K.}~\bibnamefont
  {Tanigaki}}, \bibinfo {author} {\bibfnamefont {T.~W.}\ \bibnamefont
  {Ebbesen}}, \bibinfo {author} {\bibfnamefont {S.}~\bibnamefont {Saito}},
  \bibinfo {author} {\bibfnamefont {J.}~\bibnamefont {Mizuki}}, \bibinfo
  {author} {\bibfnamefont {J.~S.}\ \bibnamefont {Tsai}}, \bibinfo {author}
  {\bibfnamefont {Y.}~\bibnamefont {Kubo}},\ and\ \bibinfo {author}
  {\bibfnamefont {S.}~\bibnamefont {Kuroshima}},\ }\bibfield  {title} {\bibinfo
  {title} {Superconductivity at 33 k in {CsxRbyC}60},\ }\href
  {https://doi.org/10.1038/352222a0} {\bibfield  {journal} {\bibinfo  {journal}
  {Nature}\ }\textbf {\bibinfo {volume} {352}},\ \bibinfo {pages} {222}
  (\bibinfo {year} {1991})}\BibitemShut {NoStop}%
\bibitem [{\citenamefont {Tanigaki}\ \emph {et~al.}(1993)\citenamefont
  {Tanigaki}, \citenamefont {Hirosawa}, \citenamefont {Ebbesen}, \citenamefont
  {Mizuki},\ and\ \citenamefont {Tsai}}]{Tanigaki1993}%
  \BibitemOpen
  \bibfield  {author} {\bibinfo {author} {\bibfnamefont {K.}~\bibnamefont
  {Tanigaki}}, \bibinfo {author} {\bibfnamefont {I.}~\bibnamefont {Hirosawa}},
  \bibinfo {author} {\bibfnamefont {T.~W.}\ \bibnamefont {Ebbesen}}, \bibinfo
  {author} {\bibfnamefont {J.-I.}\ \bibnamefont {Mizuki}},\ and\ \bibinfo
  {author} {\bibfnamefont {J.-S.}\ \bibnamefont {Tsai}},\ }\bibfield  {title}
  {\bibinfo {title} {Structure and superconductivity of c60 fullerides},\
  }\href {https://doi.org/10.1016/0022-3697(93)90278-y} {\bibfield  {journal}
  {\bibinfo  {journal} {J. Phys. Chem. Solids}\ }\textbf {\bibinfo {volume}
  {54}},\ \bibinfo {pages} {1645} (\bibinfo {year} {1993})}\BibitemShut
  {NoStop}%
\bibitem [{\citenamefont {Palstra}\ \emph {et~al.}(1995)\citenamefont
  {Palstra}, \citenamefont {Zhou}, \citenamefont {Iwasa}, \citenamefont
  {Sulewski}, \citenamefont {Fleming},\ and\ \citenamefont
  {Zegarski}}]{Palstra1995}%
  \BibitemOpen
  \bibfield  {author} {\bibinfo {author} {\bibfnamefont {T.}~\bibnamefont
  {Palstra}}, \bibinfo {author} {\bibfnamefont {O.}~\bibnamefont {Zhou}},
  \bibinfo {author} {\bibfnamefont {Y.}~\bibnamefont {Iwasa}}, \bibinfo
  {author} {\bibfnamefont {P.}~\bibnamefont {Sulewski}}, \bibinfo {author}
  {\bibfnamefont {R.}~\bibnamefont {Fleming}},\ and\ \bibinfo {author}
  {\bibfnamefont {B.}~\bibnamefont {Zegarski}},\ }\bibfield  {title} {\bibinfo
  {title} {Superconductivity at 40k in cesium doped c60},\ }\href
  {https://doi.org/10.1016/0038-1098(94)00787-x} {\bibfield  {journal}
  {\bibinfo  {journal} {Solid State Commun.}\ }\textbf {\bibinfo {volume}
  {93}},\ \bibinfo {pages} {327} (\bibinfo {year} {1995})}\BibitemShut
  {NoStop}%
\bibitem [{\citenamefont {Tang}(2001)}]{Tang2001}%
  \BibitemOpen
  \bibfield  {author} {\bibinfo {author} {\bibfnamefont {Z.~K.}\ \bibnamefont
  {Tang}},\ }\bibfield  {title} {\bibinfo {title} {Superconductivity in 4
  angstrom single-walled carbon nanotubes},\ }\href
  {https://doi.org/10.1126/science.1060470} {\bibfield  {journal} {\bibinfo
  {journal} {Science}\ }\textbf {\bibinfo {volume} {292}},\ \bibinfo {pages}
  {2462} (\bibinfo {year} {2001})}\BibitemShut {NoStop}%
\bibitem [{\citenamefont {Cao}\ \emph {et~al.}(2018{\natexlab{a}})\citenamefont
  {Cao}, \citenamefont {Fatemi}, \citenamefont {Demir}, \citenamefont {Fang},
  \citenamefont {Tomarken}, \citenamefont {Luo}, \citenamefont
  {Sanchez-Yamagishi}, \citenamefont {Watanabe}, \citenamefont {Taniguchi},
  \citenamefont {Kaxiras}, \citenamefont {Ashoori},\ and\ \citenamefont
  {Jarillo-Herrero}}]{Cao2018}%
  \BibitemOpen
  \bibfield  {author} {\bibinfo {author} {\bibfnamefont {Y.}~\bibnamefont
  {Cao}}, \bibinfo {author} {\bibfnamefont {V.}~\bibnamefont {Fatemi}},
  \bibinfo {author} {\bibfnamefont {A.}~\bibnamefont {Demir}}, \bibinfo
  {author} {\bibfnamefont {S.}~\bibnamefont {Fang}}, \bibinfo {author}
  {\bibfnamefont {S.~L.}\ \bibnamefont {Tomarken}}, \bibinfo {author}
  {\bibfnamefont {J.~Y.}\ \bibnamefont {Luo}}, \bibinfo {author} {\bibfnamefont
  {J.~D.}\ \bibnamefont {Sanchez-Yamagishi}}, \bibinfo {author} {\bibfnamefont
  {K.}~\bibnamefont {Watanabe}}, \bibinfo {author} {\bibfnamefont
  {T.}~\bibnamefont {Taniguchi}}, \bibinfo {author} {\bibfnamefont
  {E.}~\bibnamefont {Kaxiras}}, \bibinfo {author} {\bibfnamefont {R.~C.}\
  \bibnamefont {Ashoori}},\ and\ \bibinfo {author} {\bibfnamefont
  {P.}~\bibnamefont {Jarillo-Herrero}},\ }\bibfield  {title} {\bibinfo {title}
  {Correlated insulator behaviour at half-filling in magic-angle graphene
  superlattices},\ }\href {https://doi.org/10.1038/nature26154} {\bibfield
  {journal} {\bibinfo  {journal} {Nature}\ }\textbf {\bibinfo {volume} {556}},\
  \bibinfo {pages} {80} (\bibinfo {year} {2018}{\natexlab{a}})}\BibitemShut
  {NoStop}%
\bibitem [{\citenamefont {Cao}\ \emph {et~al.}(2018{\natexlab{b}})\citenamefont
  {Cao}, \citenamefont {Fatemi}, \citenamefont {Fang}, \citenamefont
  {Watanabe}, \citenamefont {Taniguchi}, \citenamefont {Kaxiras},\ and\
  \citenamefont {Jarillo-Herrero}}]{Cao2018a}%
  \BibitemOpen
  \bibfield  {author} {\bibinfo {author} {\bibfnamefont {Y.}~\bibnamefont
  {Cao}}, \bibinfo {author} {\bibfnamefont {V.}~\bibnamefont {Fatemi}},
  \bibinfo {author} {\bibfnamefont {S.}~\bibnamefont {Fang}}, \bibinfo {author}
  {\bibfnamefont {K.}~\bibnamefont {Watanabe}}, \bibinfo {author}
  {\bibfnamefont {T.}~\bibnamefont {Taniguchi}}, \bibinfo {author}
  {\bibfnamefont {E.}~\bibnamefont {Kaxiras}},\ and\ \bibinfo {author}
  {\bibfnamefont {P.}~\bibnamefont {Jarillo-Herrero}},\ }\bibfield  {title}
  {\bibinfo {title} {Unconventional superconductivity in magic-angle graphene
  superlattices},\ }\href {https://doi.org/10.1038/nature26160} {\bibfield
  {journal} {\bibinfo  {journal} {Nature}\ }\textbf {\bibinfo {volume} {556}},\
  \bibinfo {pages} {43} (\bibinfo {year} {2018}{\natexlab{b}})}\BibitemShut
  {NoStop}%
\bibitem [{\citenamefont {Sheng}\ \emph {et~al.}(2011)\citenamefont {Sheng},
  \citenamefont {Yan}, \citenamefont {Ye}, \citenamefont {Zheng},\ and\
  \citenamefont {Su}}]{Sheng2011}%
  \BibitemOpen
  \bibfield  {author} {\bibinfo {author} {\bibfnamefont {X.-L.}\ \bibnamefont
  {Sheng}}, \bibinfo {author} {\bibfnamefont {Q.-B.}\ \bibnamefont {Yan}},
  \bibinfo {author} {\bibfnamefont {F.}~\bibnamefont {Ye}}, \bibinfo {author}
  {\bibfnamefont {Q.-R.}\ \bibnamefont {Zheng}},\ and\ \bibinfo {author}
  {\bibfnamefont {G.}~\bibnamefont {Su}},\ }\bibfield  {title} {\bibinfo
  {title} {T-carbon: A novel carbon allotrope},\ }\href
  {https://doi.org/10.1103/physrevlett.106.155703} {\bibfield  {journal}
  {\bibinfo  {journal} {Phys. Rev. Lett.}\ }\textbf {\bibinfo {volume} {106}},\
  \bibinfo {pages} {155703} (\bibinfo {year} {2011})}\BibitemShut {NoStop}%
\bibitem [{\citenamefont {Zhang}\ \emph {et~al.}(2017)\citenamefont {Zhang},
  \citenamefont {Wang}, \citenamefont {Zhu}, \citenamefont {Pan}, \citenamefont
  {Han}, \citenamefont {Li}, \citenamefont {Zhao}, \citenamefont {Ma},
  \citenamefont {Wang}, \citenamefont {Su},\ and\ \citenamefont
  {Niu}}]{Zhang2017}%
  \BibitemOpen
  \bibfield  {author} {\bibinfo {author} {\bibfnamefont {J.}~\bibnamefont
  {Zhang}}, \bibinfo {author} {\bibfnamefont {R.}~\bibnamefont {Wang}},
  \bibinfo {author} {\bibfnamefont {X.}~\bibnamefont {Zhu}}, \bibinfo {author}
  {\bibfnamefont {A.}~\bibnamefont {Pan}}, \bibinfo {author} {\bibfnamefont
  {C.}~\bibnamefont {Han}}, \bibinfo {author} {\bibfnamefont {X.}~\bibnamefont
  {Li}}, \bibinfo {author} {\bibfnamefont {D.}~\bibnamefont {Zhao}}, \bibinfo
  {author} {\bibfnamefont {C.}~\bibnamefont {Ma}}, \bibinfo {author}
  {\bibfnamefont {W.}~\bibnamefont {Wang}}, \bibinfo {author} {\bibfnamefont
  {H.}~\bibnamefont {Su}},\ and\ \bibinfo {author} {\bibfnamefont
  {C.}~\bibnamefont {Niu}},\ }\bibfield  {title} {\bibinfo {title}
  {Pseudo-topotactic conversion of carbon nanotubes to t-carbon nanowires under
  picosecond laser irradiation in methanol},\ }\href
  {https://doi.org/10.1038/s41467-017-00817-9} {\bibfield  {journal} {\bibinfo
  {journal} {Nat. Commun.}\ }\textbf {\bibinfo {volume} {8}},\ \bibinfo {pages}
  {683} (\bibinfo {year} {2017})}\BibitemShut {NoStop}%
\bibitem [{\citenamefont {Xu}\ \emph {et~al.}(2020)\citenamefont {Xu},
  \citenamefont {Liu}, \citenamefont {Shi}, \citenamefont {You}, \citenamefont
  {Ma}, \citenamefont {Cui}, \citenamefont {Yan}, \citenamefont {Chen},\ and\
  \citenamefont {Su}}]{Xu2020}%
  \BibitemOpen
  \bibfield  {author} {\bibinfo {author} {\bibfnamefont {K.}~\bibnamefont
  {Xu}}, \bibinfo {author} {\bibfnamefont {H.}~\bibnamefont {Liu}}, \bibinfo
  {author} {\bibfnamefont {Y.-C.}\ \bibnamefont {Shi}}, \bibinfo {author}
  {\bibfnamefont {J.-Y.}\ \bibnamefont {You}}, \bibinfo {author} {\bibfnamefont
  {X.-Y.}\ \bibnamefont {Ma}}, \bibinfo {author} {\bibfnamefont {H.-J.}\
  \bibnamefont {Cui}}, \bibinfo {author} {\bibfnamefont {Q.-B.}\ \bibnamefont
  {Yan}}, \bibinfo {author} {\bibfnamefont {G.-C.}\ \bibnamefont {Chen}},\ and\
  \bibinfo {author} {\bibfnamefont {G.}~\bibnamefont {Su}},\ }\bibfield
  {title} {\bibinfo {title} {Preparation of t-carbon by plasma enhanced
  chemical vapor deposition},\ }\href
  {https://doi.org/10.1016/j.carbon.2019.10.032} {\bibfield  {journal}
  {\bibinfo  {journal} {Carbon}\ }\textbf {\bibinfo {volume} {157}},\ \bibinfo
  {pages} {270} (\bibinfo {year} {2020})}\BibitemShut {NoStop}%
\bibitem [{\citenamefont {Qin}\ \emph {et~al.}(2019)\citenamefont {Qin},
  \citenamefont {Hao}, \citenamefont {Yan}, \citenamefont {Hu},\ and\
  \citenamefont {Su}}]{Qin2019}%
  \BibitemOpen
  \bibfield  {author} {\bibinfo {author} {\bibfnamefont {G.}~\bibnamefont
  {Qin}}, \bibinfo {author} {\bibfnamefont {K.-R.}\ \bibnamefont {Hao}},
  \bibinfo {author} {\bibfnamefont {Q.-B.}\ \bibnamefont {Yan}}, \bibinfo
  {author} {\bibfnamefont {M.}~\bibnamefont {Hu}},\ and\ \bibinfo {author}
  {\bibfnamefont {G.}~\bibnamefont {Su}},\ }\bibfield  {title} {\bibinfo
  {title} {Exploring t-carbon for energy applications},\ }\href
  {https://doi.org/10.1039/c8nr09557d} {\bibfield  {journal} {\bibinfo
  {journal} {Nanoscale}\ }\textbf {\bibinfo {volume} {11}},\ \bibinfo {pages}
  {5798} (\bibinfo {year} {2019})}\BibitemShut {NoStop}%
\bibitem [{\citenamefont {Bai}\ \emph {et~al.}(2018)\citenamefont {Bai},
  \citenamefont {Sun}, \citenamefont {Liu}, \citenamefont {Liu},\ and\
  \citenamefont {Zhou}}]{Bai2018}%
  \BibitemOpen
  \bibfield  {author} {\bibinfo {author} {\bibfnamefont {L.}~\bibnamefont
  {Bai}}, \bibinfo {author} {\bibfnamefont {P.-P.}\ \bibnamefont {Sun}},
  \bibinfo {author} {\bibfnamefont {B.}~\bibnamefont {Liu}}, \bibinfo {author}
  {\bibfnamefont {Z.}~\bibnamefont {Liu}},\ and\ \bibinfo {author}
  {\bibfnamefont {K.}~\bibnamefont {Zhou}},\ }\bibfield  {title} {\bibinfo
  {title} {Mechanical behaviors of t-carbon: A molecular dynamics study},\
  }\href {https://doi.org/10.1016/j.carbon.2018.07.046} {\bibfield  {journal}
  {\bibinfo  {journal} {Carbon}\ }\textbf {\bibinfo {volume} {138}},\ \bibinfo
  {pages} {357} (\bibinfo {year} {2018})}\BibitemShut {NoStop}%
\bibitem [{\citenamefont {Wang}\ \emph {et~al.}(2019)\citenamefont {Wang},
  \citenamefont {Lei}, \citenamefont {Bai}, \citenamefont {Zhou},\ and\
  \citenamefont {Liu}}]{Wang2019}%
  \BibitemOpen
  \bibfield  {author} {\bibinfo {author} {\bibfnamefont {Y.}~\bibnamefont
  {Wang}}, \bibinfo {author} {\bibfnamefont {J.}~\bibnamefont {Lei}}, \bibinfo
  {author} {\bibfnamefont {L.}~\bibnamefont {Bai}}, \bibinfo {author}
  {\bibfnamefont {K.}~\bibnamefont {Zhou}},\ and\ \bibinfo {author}
  {\bibfnamefont {Z.}~\bibnamefont {Liu}},\ }\bibfield  {title} {\bibinfo
  {title} {Effects of tensile strain rate and grain size on the mechanical
  properties of nanocrystalline t-carbon},\ }\href
  {https://doi.org/10.1016/j.commatsci.2019.109188} {\bibfield  {journal}
  {\bibinfo  {journal} {Computational Materials Science}\ }\textbf {\bibinfo
  {volume} {170}},\ \bibinfo {pages} {109188} (\bibinfo {year}
  {2019})}\BibitemShut {NoStop}%
\bibitem [{\citenamefont {Yue}\ \emph {et~al.}(2017)\citenamefont {Yue},
  \citenamefont {Qin}, \citenamefont {Zhang}, \citenamefont {Sheng},
  \citenamefont {Su},\ and\ \citenamefont {Hu}}]{Yue2017}%
  \BibitemOpen
  \bibfield  {author} {\bibinfo {author} {\bibfnamefont {S.-Y.}\ \bibnamefont
  {Yue}}, \bibinfo {author} {\bibfnamefont {G.}~\bibnamefont {Qin}}, \bibinfo
  {author} {\bibfnamefont {X.}~\bibnamefont {Zhang}}, \bibinfo {author}
  {\bibfnamefont {X.}~\bibnamefont {Sheng}}, \bibinfo {author} {\bibfnamefont
  {G.}~\bibnamefont {Su}},\ and\ \bibinfo {author} {\bibfnamefont
  {M.}~\bibnamefont {Hu}},\ }\bibfield  {title} {\bibinfo {title} {Thermal
  transport in novel carbon allotropes with sp2 or sp3 hybridization: An ab
  initio study},\ }\href {https://doi.org/10.1103/physrevb.95.085207}
  {\bibfield  {journal} {\bibinfo  {journal} {Phys. Rev. B}\ }\textbf {\bibinfo
  {volume} {95}},\ \bibinfo {pages} {085207} (\bibinfo {year}
  {2017})}\BibitemShut {NoStop}%
\bibitem [{\citenamefont {Ren}\ \emph {et~al.}(2019)\citenamefont {Ren},
  \citenamefont {Chu}, \citenamefont {Li}, \citenamefont {Yue},\ and\
  \citenamefont {Hu}}]{Ren2019}%
  \BibitemOpen
  \bibfield  {author} {\bibinfo {author} {\bibfnamefont {H.}~\bibnamefont
  {Ren}}, \bibinfo {author} {\bibfnamefont {H.}~\bibnamefont {Chu}}, \bibinfo
  {author} {\bibfnamefont {Z.}~\bibnamefont {Li}}, \bibinfo {author}
  {\bibfnamefont {T.}~\bibnamefont {Yue}},\ and\ \bibinfo {author}
  {\bibfnamefont {Z.}~\bibnamefont {Hu}},\ }\bibfield  {title} {\bibinfo
  {title} {Efficient energy gap tuning for t-carbon via single atomic doping},\
  }\href {https://doi.org/10.1016/j.chemphys.2018.11.005} {\bibfield  {journal}
  {\bibinfo  {journal} {Chem. Phys.}\ }\textbf {\bibinfo {volume} {518}},\
  \bibinfo {pages} {69} (\bibinfo {year} {2019})}\BibitemShut {NoStop}%
\bibitem [{\citenamefont {Alborznia}\ \emph {et~al.}(2019)\citenamefont
  {Alborznia}, \citenamefont {Naseri},\ and\ \citenamefont
  {Fatahi}}]{Alborznia2019}%
  \BibitemOpen
  \bibfield  {author} {\bibinfo {author} {\bibfnamefont {H.}~\bibnamefont
  {Alborznia}}, \bibinfo {author} {\bibfnamefont {M.}~\bibnamefont {Naseri}},\
  and\ \bibinfo {author} {\bibfnamefont {N.}~\bibnamefont {Fatahi}},\
  }\bibfield  {title} {\bibinfo {title} {Pressure effects on the optical and
  electronic aspects of t-carbon: A first principles calculation},\ }\href
  {https://doi.org/10.1016/j.ijleo.2018.11.077} {\bibfield  {journal} {\bibinfo
   {journal} {Optik}\ }\textbf {\bibinfo {volume} {180}},\ \bibinfo {pages}
  {125} (\bibinfo {year} {2019})}\BibitemShut {NoStop}%
\bibitem [{\citenamefont {Sun}\ \emph {et~al.}(2019)\citenamefont {Sun},
  \citenamefont {Bai}, \citenamefont {Kripalani},\ and\ \citenamefont
  {Zhou}}]{Sun2019}%
  \BibitemOpen
  \bibfield  {author} {\bibinfo {author} {\bibfnamefont {P.-P.}\ \bibnamefont
  {Sun}}, \bibinfo {author} {\bibfnamefont {L.}~\bibnamefont {Bai}}, \bibinfo
  {author} {\bibfnamefont {D.~R.}\ \bibnamefont {Kripalani}},\ and\ \bibinfo
  {author} {\bibfnamefont {K.}~\bibnamefont {Zhou}},\ }\bibfield  {title}
  {\bibinfo {title} {A new carbon phase with direct bandgap and high carrier
  mobility as electron transport material for perovskite solar cells},\ }\href
  {https://doi.org/10.1038/s41524-018-0146-z} {\bibfield  {journal} {\bibinfo
  {journal} {npj Comput. Mater.}\ }\textbf {\bibinfo {volume} {5}},\ \bibinfo
  {pages} {9} (\bibinfo {year} {2019})}\BibitemShut {NoStop}%
\bibitem [{\citenamefont {Gharsallah}\ \emph {et~al.}(2016)\citenamefont
  {Gharsallah}, \citenamefont {Serrano-S{\'{a}}nchez}, \citenamefont {Nemes},
  \citenamefont {Mompe{\'{a}}n}, \citenamefont {Mart{\'{\i}}nez}, \citenamefont
  {Fern{\'{a}}ndez-D{\'{\i}}az}, \citenamefont {Elhalouani},\ and\
  \citenamefont {Alonso}}]{Gharsallah2016}%
  \BibitemOpen
  \bibfield  {author} {\bibinfo {author} {\bibfnamefont {M.}~\bibnamefont
  {Gharsallah}}, \bibinfo {author} {\bibfnamefont {F.}~\bibnamefont
  {Serrano-S{\'{a}}nchez}}, \bibinfo {author} {\bibfnamefont {N.~M.}\
  \bibnamefont {Nemes}}, \bibinfo {author} {\bibfnamefont {F.~J.}\ \bibnamefont
  {Mompe{\'{a}}n}}, \bibinfo {author} {\bibfnamefont {J.~L.}\ \bibnamefont
  {Mart{\'{\i}}nez}}, \bibinfo {author} {\bibfnamefont {M.~T.}\ \bibnamefont
  {Fern{\'{a}}ndez-D{\'{\i}}az}}, \bibinfo {author} {\bibfnamefont
  {F.}~\bibnamefont {Elhalouani}},\ and\ \bibinfo {author} {\bibfnamefont
  {J.~A.}\ \bibnamefont {Alonso}},\ }\bibfield  {title} {\bibinfo {title}
  {Giant seebeck effect in ge-doped {SnSe}},\ }\href
  {https://doi.org/10.1038/srep26774} {\bibfield  {journal} {\bibinfo
  {journal} {Sci. Rep.}\ }\textbf {\bibinfo {volume} {6}},\ \bibinfo {pages}
  {26774} (\bibinfo {year} {2016})}\BibitemShut {NoStop}%
\bibitem [{\citenamefont {Qin}\ \emph {et~al.}(2016)\citenamefont {Qin},
  \citenamefont {Qin}, \citenamefont {Fang}, \citenamefont {Zhang},
  \citenamefont {Yue}, \citenamefont {Yan}, \citenamefont {Hu},\ and\
  \citenamefont {Su}}]{Qin2016}%
  \BibitemOpen
  \bibfield  {author} {\bibinfo {author} {\bibfnamefont {G.}~\bibnamefont
  {Qin}}, \bibinfo {author} {\bibfnamefont {Z.}~\bibnamefont {Qin}}, \bibinfo
  {author} {\bibfnamefont {W.-Z.}\ \bibnamefont {Fang}}, \bibinfo {author}
  {\bibfnamefont {L.-C.}\ \bibnamefont {Zhang}}, \bibinfo {author}
  {\bibfnamefont {S.-Y.}\ \bibnamefont {Yue}}, \bibinfo {author} {\bibfnamefont
  {Q.-B.}\ \bibnamefont {Yan}}, \bibinfo {author} {\bibfnamefont
  {M.}~\bibnamefont {Hu}},\ and\ \bibinfo {author} {\bibfnamefont
  {G.}~\bibnamefont {Su}},\ }\bibfield  {title} {\bibinfo {title} {Diverse
  anisotropy of phonon transport in two-dimensional group {IV}{\textendash}{VI}
  compounds: A comparative study},\ }\href {https://doi.org/10.1039/c6nr01349j}
  {\bibfield  {journal} {\bibinfo  {journal} {Nanoscale}\ }\textbf {\bibinfo
  {volume} {8}},\ \bibinfo {pages} {11306} (\bibinfo {year}
  {2016})}\BibitemShut {NoStop}%
\bibitem [{\citenamefont {Giannozzi}\ \emph {et~al.}(2009)\citenamefont
  {Giannozzi}, \citenamefont {Baroni}, \citenamefont {Bonini}, \citenamefont
  {Calandra}, \citenamefont {Car}, \citenamefont {Cavazzoni}, \citenamefont
  {Ceresoli}, \citenamefont {Chiarotti}, \citenamefont {Cococcioni},
  \citenamefont {Dabo}, \citenamefont {Corso}, \citenamefont {de~Gironcoli},
  \citenamefont {Fabris}, \citenamefont {Fratesi}, \citenamefont {Gebauer},
  \citenamefont {Gerstmann}, \citenamefont {Gougoussis}, \citenamefont
  {Kokalj}, \citenamefont {Lazzeri}, \citenamefont {Martin-Samos},
  \citenamefont {Marzari}, \citenamefont {Mauri}, \citenamefont {Mazzarello},
  \citenamefont {Paolini}, \citenamefont {Pasquarello}, \citenamefont
  {Paulatto}, \citenamefont {Sbraccia}, \citenamefont {Scandolo}, \citenamefont
  {Sclauzero}, \citenamefont {Seitsonen}, \citenamefont {Smogunov},
  \citenamefont {Umari},\ and\ \citenamefont {Wentzcovitch}}]{Giannozzi2009}%
  \BibitemOpen
  \bibfield  {author} {\bibinfo {author} {\bibfnamefont {P.}~\bibnamefont
  {Giannozzi}}, \bibinfo {author} {\bibfnamefont {S.}~\bibnamefont {Baroni}},
  \bibinfo {author} {\bibfnamefont {N.}~\bibnamefont {Bonini}}, \bibinfo
  {author} {\bibfnamefont {M.}~\bibnamefont {Calandra}}, \bibinfo {author}
  {\bibfnamefont {R.}~\bibnamefont {Car}}, \bibinfo {author} {\bibfnamefont
  {C.}~\bibnamefont {Cavazzoni}}, \bibinfo {author} {\bibfnamefont
  {D.}~\bibnamefont {Ceresoli}}, \bibinfo {author} {\bibfnamefont {G.~L.}\
  \bibnamefont {Chiarotti}}, \bibinfo {author} {\bibfnamefont {M.}~\bibnamefont
  {Cococcioni}}, \bibinfo {author} {\bibfnamefont {I.}~\bibnamefont {Dabo}},
  \bibinfo {author} {\bibfnamefont {A.~D.}\ \bibnamefont {Corso}}, \bibinfo
  {author} {\bibfnamefont {S.}~\bibnamefont {de~Gironcoli}}, \bibinfo {author}
  {\bibfnamefont {S.}~\bibnamefont {Fabris}}, \bibinfo {author} {\bibfnamefont
  {G.}~\bibnamefont {Fratesi}}, \bibinfo {author} {\bibfnamefont
  {R.}~\bibnamefont {Gebauer}}, \bibinfo {author} {\bibfnamefont
  {U.}~\bibnamefont {Gerstmann}}, \bibinfo {author} {\bibfnamefont
  {C.}~\bibnamefont {Gougoussis}}, \bibinfo {author} {\bibfnamefont
  {A.}~\bibnamefont {Kokalj}}, \bibinfo {author} {\bibfnamefont
  {M.}~\bibnamefont {Lazzeri}}, \bibinfo {author} {\bibfnamefont
  {L.}~\bibnamefont {Martin-Samos}}, \bibinfo {author} {\bibfnamefont
  {N.}~\bibnamefont {Marzari}}, \bibinfo {author} {\bibfnamefont
  {F.}~\bibnamefont {Mauri}}, \bibinfo {author} {\bibfnamefont
  {R.}~\bibnamefont {Mazzarello}}, \bibinfo {author} {\bibfnamefont
  {S.}~\bibnamefont {Paolini}}, \bibinfo {author} {\bibfnamefont
  {A.}~\bibnamefont {Pasquarello}}, \bibinfo {author} {\bibfnamefont
  {L.}~\bibnamefont {Paulatto}}, \bibinfo {author} {\bibfnamefont
  {C.}~\bibnamefont {Sbraccia}}, \bibinfo {author} {\bibfnamefont
  {S.}~\bibnamefont {Scandolo}}, \bibinfo {author} {\bibfnamefont
  {G.}~\bibnamefont {Sclauzero}}, \bibinfo {author} {\bibfnamefont {A.~P.}\
  \bibnamefont {Seitsonen}}, \bibinfo {author} {\bibfnamefont {A.}~\bibnamefont
  {Smogunov}}, \bibinfo {author} {\bibfnamefont {P.}~\bibnamefont {Umari}},\
  and\ \bibinfo {author} {\bibfnamefont {R.~M.}\ \bibnamefont {Wentzcovitch}},\
  }\bibfield  {title} {\bibinfo {title} {{QUANTUM} {ESPRESSO}: a modular and
  open-source software project for quantum simulations of materials},\ }\href
  {https://doi.org/10.1088/0953-8984/21/39/395502} {\bibfield  {journal}
  {\bibinfo  {journal} {J. Phys.: Condens. Matter}\ }\textbf {\bibinfo {volume}
  {21}},\ \bibinfo {pages} {395502} (\bibinfo {year} {2009})}\BibitemShut
  {NoStop}%
\bibitem [{\citenamefont {Blöchl}(1994)}]{Bloechl1994}%
  \BibitemOpen
  \bibfield  {author} {\bibinfo {author} {\bibfnamefont {P.~E.}\ \bibnamefont
  {Blöchl}},\ }\bibfield  {title} {\bibinfo {title} {Projector augmented-wave
  method},\ }\href {https://doi.org/10.1103/physrevb.50.17953} {\bibfield
  {journal} {\bibinfo  {journal} {Phys. Rev. B}\ }\textbf {\bibinfo {volume}
  {50}},\ \bibinfo {pages} {17953} (\bibinfo {year} {1994})}\BibitemShut
  {NoStop}%
\bibitem [{\citenamefont {Perdew}\ \emph {et~al.}(1996)\citenamefont {Perdew},
  \citenamefont {Burke},\ and\ \citenamefont {Ernzerhof}}]{Perdew1996}%
  \BibitemOpen
  \bibfield  {author} {\bibinfo {author} {\bibfnamefont {J.~P.}\ \bibnamefont
  {Perdew}}, \bibinfo {author} {\bibfnamefont {K.}~\bibnamefont {Burke}},\ and\
  \bibinfo {author} {\bibfnamefont {M.}~\bibnamefont {Ernzerhof}},\ }\bibfield
  {title} {\bibinfo {title} {Generalized gradient approximation made simple},\
  }\href {https://doi.org/10.1103/physrevlett.77.3865} {\bibfield  {journal}
  {\bibinfo  {journal} {Phys. Rev. Lett.}\ }\textbf {\bibinfo {volume} {77}},\
  \bibinfo {pages} {3865} (\bibinfo {year} {1996})}\BibitemShut {NoStop}%
\bibitem [{\citenamefont {Baroni}\ \emph {et~al.}(2001)\citenamefont {Baroni},
  \citenamefont {de~Gironcoli}, \citenamefont {Corso},\ and\ \citenamefont
  {Giannozzi}}]{Baroni2001}%
  \BibitemOpen
  \bibfield  {author} {\bibinfo {author} {\bibfnamefont {S.}~\bibnamefont
  {Baroni}}, \bibinfo {author} {\bibfnamefont {S.}~\bibnamefont
  {de~Gironcoli}}, \bibinfo {author} {\bibfnamefont {A.~D.}\ \bibnamefont
  {Corso}},\ and\ \bibinfo {author} {\bibfnamefont {P.}~\bibnamefont
  {Giannozzi}},\ }\bibfield  {title} {\bibinfo {title} {Phonons and related
  crystal properties from density-functional perturbation theory},\ }\href
  {https://doi.org/10.1103/revmodphys.73.515} {\bibfield  {journal} {\bibinfo
  {journal} {Rev. Mod. Phys.}\ }\textbf {\bibinfo {volume} {73}},\ \bibinfo
  {pages} {515} (\bibinfo {year} {2001})}\BibitemShut {NoStop}%
\bibitem [{\citenamefont {Madsen}\ and\ \citenamefont
  {Singh}(2006)}]{Madsen2006}%
  \BibitemOpen
  \bibfield  {author} {\bibinfo {author} {\bibfnamefont {G.~K.}\ \bibnamefont
  {Madsen}}\ and\ \bibinfo {author} {\bibfnamefont {D.~J.}\ \bibnamefont
  {Singh}},\ }\bibfield  {title} {\bibinfo {title} {{BoltzTraP}. a code for
  calculating band-structure dependent quantities},\ }\href
  {https://doi.org/10.1016/j.cpc.2006.03.007} {\bibfield  {journal} {\bibinfo
  {journal} {Comput. Phys. Commun.}\ }\textbf {\bibinfo {volume} {175}},\
  \bibinfo {pages} {67} (\bibinfo {year} {2006})}\BibitemShut {NoStop}%
\bibitem [{\citenamefont {Zhu}\ \emph {et~al.}(2018)\citenamefont {Zhu},
  \citenamefont {Liu}, \citenamefont {Yu}, \citenamefont {Wang}, \citenamefont
  {Zhao}, \citenamefont {Feng}, \citenamefont {Sheng},\ and\ \citenamefont
  {Yang}}]{Zhu2018}%
  \BibitemOpen
  \bibfield  {author} {\bibinfo {author} {\bibfnamefont {Z.}~\bibnamefont
  {Zhu}}, \bibinfo {author} {\bibfnamefont {Y.}~\bibnamefont {Liu}}, \bibinfo
  {author} {\bibfnamefont {Z.-M.}\ \bibnamefont {Yu}}, \bibinfo {author}
  {\bibfnamefont {S.-S.}\ \bibnamefont {Wang}}, \bibinfo {author}
  {\bibfnamefont {Y.~X.}\ \bibnamefont {Zhao}}, \bibinfo {author}
  {\bibfnamefont {Y.}~\bibnamefont {Feng}}, \bibinfo {author} {\bibfnamefont
  {X.-L.}\ \bibnamefont {Sheng}},\ and\ \bibinfo {author} {\bibfnamefont
  {S.~A.}\ \bibnamefont {Yang}},\ }\bibfield  {title} {\bibinfo {title}
  {Quadratic contact point semimetal: Theory and material realization},\ }\href
  {https://doi.org/10.1103/physrevb.98.125104} {\bibfield  {journal} {\bibinfo
  {journal} {Phys. Rev. B}\ }\textbf {\bibinfo {volume} {98}},\ \bibinfo
  {pages} {125104} (\bibinfo {year} {2018})}\BibitemShut {NoStop}%
\bibitem [{\citenamefont {Güntherodt}(1980)}]{Guentherodt1980}%
  \BibitemOpen
  \bibfield  {author} {\bibinfo {author} {\bibfnamefont {G.}~\bibnamefont
  {Güntherodt}},\ }\bibfield  {title} {\bibinfo {title} {{ELECTRON}-{PHONON}
  {INTERACTION} {IN} {SmS}},\ }\href {https://doi.org/10.1051/jphyscol:1980512}
  {\bibfield  {journal} {\bibinfo  {journal} {Le Journal de Physique
  Colloques}\ }\textbf {\bibinfo {volume} {41}},\ \bibinfo {pages} {C5}
  (\bibinfo {year} {1980})}\BibitemShut {NoStop}%
\bibitem [{\citenamefont {Giustino}(2017)}]{Giustino2017}%
  \BibitemOpen
  \bibfield  {author} {\bibinfo {author} {\bibfnamefont {F.}~\bibnamefont
  {Giustino}},\ }\bibfield  {title} {\bibinfo {title} {Electron-phonon
  interactions from first principles},\ }\href
  {https://doi.org/10.1103/revmodphys.89.015003} {\bibfield  {journal}
  {\bibinfo  {journal} {Rev. Mod. Phys.}\ }\textbf {\bibinfo {volume} {89}},\
  \bibinfo {pages} {015003} (\bibinfo {year} {2017})}\BibitemShut {NoStop}%
\bibitem [{\citenamefont {Allen}\ and\ \citenamefont
  {Dynes}(1975)}]{Allen1975}%
  \BibitemOpen
  \bibfield  {author} {\bibinfo {author} {\bibfnamefont {P.~B.}\ \bibnamefont
  {Allen}}\ and\ \bibinfo {author} {\bibfnamefont {R.~C.}\ \bibnamefont
  {Dynes}},\ }\bibfield  {title} {\bibinfo {title} {Transition temperature of
  strong-coupled superconductors reanalyzed},\ }\href
  {https://doi.org/10.1103/physrevb.12.905} {\bibfield  {journal} {\bibinfo
  {journal} {Phys. Rev. B}\ }\textbf {\bibinfo {volume} {12}},\ \bibinfo
  {pages} {905} (\bibinfo {year} {1975})}\BibitemShut {NoStop}%
\bibitem [{\citenamefont {Nomura}\ \emph {et~al.}(2016)\citenamefont {Nomura},
  \citenamefont {Sakai}, \citenamefont {Capone},\ and\ \citenamefont
  {Arita}}]{Nomura2016}%
  \BibitemOpen
  \bibfield  {author} {\bibinfo {author} {\bibfnamefont {Y.}~\bibnamefont
  {Nomura}}, \bibinfo {author} {\bibfnamefont {S.}~\bibnamefont {Sakai}},
  \bibinfo {author} {\bibfnamefont {M.}~\bibnamefont {Capone}},\ and\ \bibinfo
  {author} {\bibfnamefont {R.}~\bibnamefont {Arita}},\ }\bibfield  {title}
  {\bibinfo {title} {Exotics-wave superconductivity in alkali-doped
  fullerides},\ }\href {https://doi.org/10.1088/0953-8984/28/15/153001}
  {\bibfield  {journal} {\bibinfo  {journal} {J. Phys.: Condens. Matter}\
  }\textbf {\bibinfo {volume} {28}},\ \bibinfo {pages} {153001} (\bibinfo
  {year} {2016})}\BibitemShut {NoStop}%
\bibitem [{\citenamefont {Schilling}()}]{Schillinga}%
  \BibitemOpen
  \bibfield  {author} {\bibinfo {author} {\bibfnamefont {J.~S.}\ \bibnamefont
  {Schilling}},\ }\bibfield  {title} {\bibinfo {title} {High-pressure
  effects},\ }in\ \href {https://doi.org/10.1007/978-0-387-68734-6_11} {\emph
  {\bibinfo {booktitle} {Handbook of High-Temperature Superconductivity}}}\
  (\bibinfo  {publisher} {Springer New York})\ pp.\ \bibinfo {pages}
  {427--462}\BibitemShut {NoStop}%
\bibitem [{\citenamefont {Hopfield}(1971)}]{Hopfield1971}%
  \BibitemOpen
  \bibfield  {author} {\bibinfo {author} {\bibfnamefont {J.}~\bibnamefont
  {Hopfield}},\ }\bibfield  {title} {\bibinfo {title} {On the systematics of
  high tc in transition metal materials},\ }\href
  {https://doi.org/10.1016/0031-8914(71)90239-4} {\bibfield  {journal}
  {\bibinfo  {journal} {Physica}\ }\textbf {\bibinfo {volume} {55}},\ \bibinfo
  {pages} {41} (\bibinfo {year} {1971})}\BibitemShut {NoStop}%
\bibitem [{\citenamefont {Chen}\ \emph {et~al.}(2002)\citenamefont {Chen},
  \citenamefont {Zhang},\ and\ \citenamefont {Habermeier}}]{Chen2002}%
  \BibitemOpen
  \bibfield  {author} {\bibinfo {author} {\bibfnamefont {X.~J.}\ \bibnamefont
  {Chen}}, \bibinfo {author} {\bibfnamefont {H.}~\bibnamefont {Zhang}},\ and\
  \bibinfo {author} {\bibfnamefont {H.-U.}\ \bibnamefont {Habermeier}},\
  }\bibfield  {title} {\bibinfo {title} {Effects of pressure on the
  superconducting properties of magnesium diboride},\ }\href
  {https://doi.org/10.1103/physrevb.65.144514} {\bibfield  {journal} {\bibinfo
  {journal} {Phys. Rev. B}\ }\textbf {\bibinfo {volume} {65}},\ \bibinfo
  {pages} {144514} (\bibinfo {year} {2002})}\BibitemShut {NoStop}%
\bibitem [{\citenamefont {Ruoff}\ and\ \citenamefont
  {Kadish}(1997)}]{ruoff1997recent}%
  \BibitemOpen
  \bibfield  {author} {\bibinfo {author} {\bibfnamefont {R.~S.}\ \bibnamefont
  {Ruoff}}\ and\ \bibinfo {author} {\bibfnamefont {K.~M.}\ \bibnamefont
  {Kadish}},\ }\bibfield  {title} {\bibinfo {title} {Recent advances in the
  chemistry and physics of fullerenes and related materials},\ }\href@noop {}
  {\bibfield  {journal} {\bibinfo  {journal} {Proceedings Volume}\ }\textbf
  {\bibinfo {volume} {95}},\ \bibinfo {pages} {10} (\bibinfo {year}
  {1997})}\BibitemShut {NoStop}%
\bibitem [{\citenamefont {Diederichs}\ \emph {et~al.}(1996)\citenamefont
  {Diederichs}, \citenamefont {Gangopadhyay},\ and\ \citenamefont
  {Schilling}}]{Diederichs1996}%
  \BibitemOpen
  \bibfield  {author} {\bibinfo {author} {\bibfnamefont {J.}~\bibnamefont
  {Diederichs}}, \bibinfo {author} {\bibfnamefont {A.~K.}\ \bibnamefont
  {Gangopadhyay}},\ and\ \bibinfo {author} {\bibfnamefont {J.~S.}\ \bibnamefont
  {Schilling}},\ }\bibfield  {title} {\bibinfo {title} {Pressure dependence of
  the electronic density of states {andTcin} {superconductingRb}3c60},\ }\href
  {https://doi.org/10.1103/physrevb.54.r9662} {\bibfield  {journal} {\bibinfo
  {journal} {Phys. Rev. B}\ }\textbf {\bibinfo {volume} {54}},\ \bibinfo
  {pages} {R9662} (\bibinfo {year} {1996})}\BibitemShut {NoStop}%
\bibitem [{\citenamefont {Diederichs}\ \emph {et~al.}(1997)\citenamefont
  {Diederichs}, \citenamefont {Schilling}, \citenamefont {Herwig},\ and\
  \citenamefont {Yelon}}]{Diederichs1997}%
  \BibitemOpen
  \bibfield  {author} {\bibinfo {author} {\bibfnamefont {J.}~\bibnamefont
  {Diederichs}}, \bibinfo {author} {\bibfnamefont {J.}~\bibnamefont
  {Schilling}}, \bibinfo {author} {\bibfnamefont {K.}~\bibnamefont {Herwig}},\
  and\ \bibinfo {author} {\bibfnamefont {W.}~\bibnamefont {Yelon}},\ }\bibfield
   {title} {\bibinfo {title} {Dependence of the superconducting transition
  temperature and lattice parameter on hydrostatic pressure for rb3c60},\
  }\href {https://doi.org/10.1016/s0022-3697(96)00087-x} {\bibfield  {journal}
  {\bibinfo  {journal} {J. Phys. Chem. Solids}\ }\textbf {\bibinfo {volume}
  {58}},\ \bibinfo {pages} {123} (\bibinfo {year} {1997})}\BibitemShut
  {NoStop}%
\end{thebibliography}

%

\end{document}